\documentclass[%
    reprint,
    superscriptaddress,
    nofootinbib,
    amsmath,amssymb,
    aps,
]{revtex4-2}

\usepackage{graphicx}
\usepackage{dcolumn}
\usepackage{bm}
\usepackage{siunitx}
\usepackage{braket}
\usepackage{comment}
\usepackage{wasysym} 
\usepackage[colorlinks=true, allcolors=blue]{hyperref}

\defcitealias{PB20}{PB20}
\defcitealias{PB22}{PB22}
\defcitealias{ALP_PB}{PB23}
\defcitealias{Takakura:2017ddx}{T17}
\defcitealias{Takakura2019}{T19}
\newcommand{\red}[1]{\textcolor{red}{#1}}
\newcommand\orcid[1]{\href{https://orcid.org/#1}{\includegraphics[height=9pt]{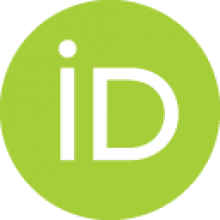}}}

\begin{document}

\preprint{APS/123-QED}

\title{Exploration of the polarization angle variability of the Crab Nebula with P\footnotesize{}OLARBEAR \\ \large{}and its application to the search for axion-like particles}

\author{Shunsuke~Adachi~\orcid{0000-0002-0400-7555}}
\affiliation{Hakubi Center for Advanced Research, Kyoto University, Kyoto 606-8501, Japan}																									
\author{Tylor~Adkins~\orcid{0000-0002-3850-9553}}
\affiliation{Department of Physics, University of California, Berkeley, CA 94720, USA}

\author{Carlo~Baccigalupi~\orcid{0000-0002-8211-1630}} 
\affiliation{Astrophysics and Cosmology, International School for Advanced Studies (SISSA), 34136 Trieste, Italy}
\affiliation{Astrophysics and Cosmology, Institute for Fundamental Physics of the Universe (IFPU), 34151 Trieste, Italy}
\affiliation{Astrophysics and Cosmology, National Institute for Nuclear Physics (INFN), 34127, Trieste, Italy}


\author{Yuji~Chinone~\orcid{0000-0002-3266-857X}}
\affiliation{QUP (WPI), KEK, Tsukuba, Ibaraki 305-0801, Japan}
\affiliation{Kavli Institute for the Physics and Mathematics of the Universe (WPI), UTIAS, The University of Tokyo, Kashiwa, Chiba 277-8583, Japan}																									
\author{Kevin~T.~Crowley~\orcid{0000-0001-5068-1295}} 
\affiliation{Department of Physics, University of California, San Diego, La Jolla, CA 92093, USA}																									
\author{Josquin~Errard~\orcid{0000-0002-1419-0031}} 
\affiliation{Universit\'{e} Paris Cit\'e, CNRS, Astroparticule et Cosmologie, F-75013 Paris, France}																									
\author{Giulio~Fabbian~\orcid{0000-0002-3255-4695}} 
\affiliation{Center for Computational Astrophysics, Flatiron Institute, New York, NY 10010, USA}
\affiliation{School of Physics and Astronomy, Cardiff University, The Parade, Cardiff, CF24 3AA, UK}																									
\author{Chang~Feng}
\affiliation{CAS Key Laboratory for Research in Galaxies and Cosmology, Department of Astronomy, University of Science and Technology of China, Hefei, Anhui 230026, China}
\affiliation{School of Astronomy and Space Science, University of Science and Technology of China, Hefei 230026, China}

\author{Takuro~Fujino~\orcid{0000-0002-1211-7850}}
\affiliation{Graduate School of Engineering Science, Yokohama National University, Yokohama 240-8501, Japan}																									
\author{Masaya~Hasegawa~\orcid{0000-0003-1443-1082}}
\affiliation{Institute of Particle and Nuclear Studies (IPNS), High Energy Accelerator Research Organization (KEK), Tsukuba, Ibaraki 305-0801, Japan}
\affiliation{QUP (WPI), KEK, Tsukuba, Ibaraki 305-0801, Japan}
\affiliation{School of High Energy Accelerator Science, The Graduate University for Advanced Studies, SOKENDAI, Kanagawa 240-0193, Japan}																									
\author{Masashi~Hazumi~\orcid{0000-0001-6830-8309}}
\affiliation{QUP (WPI), KEK, Tsukuba, Ibaraki 305-0801, Japan}
\affiliation{Institute of Particle and Nuclear Studies (IPNS), High Energy Accelerator Research Organization (KEK), Tsukuba, Ibaraki 305-0801, Japan}
\affiliation{Kavli Institute for the Physics and Mathematics of the Universe (WPI), UTIAS, The University of Tokyo, Kashiwa, Chiba 277-8583, Japan}	
\affiliation{Japan Aerospace Exploration Agency (JAXA), Institute of Space and Astronautical Science (ISAS), Sagamihara, Kanagawa 252-5210, Japan}
\affiliation{School of High Energy Accelerator Science, The Graduate University for Advanced Studies, SOKENDAI, Kanagawa 240-0193, Japan}

\author{Oliver~Jeong~\orcid{0000-0001-5893-7697}}
\affiliation{Department of Physics, University of California, Berkeley, CA 94720, USA}																									
\author{Daisuke~Kaneko~\orcid{0000-0003-3917-086X}}
\affiliation{QUP (WPI), KEK, Tsukuba, Ibaraki 305-0801, Japan}

\author{Brian~Keating~\orcid{0000-0003-3118-5514}}
\affiliation{Department of Physics, University of California, San Diego, La Jolla, CA 92093, USA}

\author{Akito~Kusaka~\orcid{0009-0004-9631-2451}}
\affiliation{Department of Physics, Graduate School of Science, The University of Tokyo, Tokyo 113-0033, Japan}
\affiliation{Physics Division, Lawrence Berkeley National Laboratory, Berkeley, CA 94720, USA}
\affiliation{Kavli Institute for the Physics and Mathematics of the Universe (WPI), UTIAS, The University of Tokyo, Kashiwa, Chiba 277-8583, Japan}
\affiliation{Research Center for the Early Universe, The University of Tokyo, Tokyo, 113-0033, Japan}

\author{Adrian~T.~Lee~\orcid{0000-0003-3106-3218}}
\affiliation{Department of Physics, University of California, Berkeley, CA 94720, USA}
\affiliation{Physics Division, Lawrence Berkeley National Laboratory, Berkeley, CA 94720, USA}

\author{Anto~I.~Lonappan~\orcid{0000-0003-1200-9179}}
\affiliation{Dipartimento di Fisica, Università di Roma “Tor Vergata”, via della Ricerca Scientifica 1, Roma I-00133, Italy}																									
\author{Yuto~Minami~\orcid{0000-0003-2176-8089}}
\affiliation{Research Center for Nuclear Physics, Osaka University, Osaka 567-0047, Japan}																									
\author{Masaaki~Murata~\orcid{0000-0003-4394-4645}}
\affiliation{Department of Physics, Graduate School of Science, The University of Tokyo, Tokyo 113-0033, Japan}																									
\author{Lucio~Piccirillo~\orcid{0000-0001-7868-0841}}
\affiliation{Jodrell Bank Centre for Astrophysics, School of Physics and Astronomy, University of
Manchester, Manchester, M13 9PL, UK}																									
\author{Christian~L.~Reichardt~\orcid{0000-0003-2226-9169}}
\affiliation{School of Physics, The University of Melbourne, Parkville VIC 3010, Australia}																									
\author{Praween~Siritanasak~\orcid{0000-0001-6830-1537}}
\affiliation{National Astronomical Research Institute of Thailand, Chiangmai, 50180, Thailand}																									
\author{Jacob~Spisak~\orcid{0000-0003-1789-8550}}
\affiliation{Department of Physics, University of California, San Diego, La Jolla, CA 92093, USA}																									
\author{Satoru~Takakura~\orcid{0000-0001-9461-7519}}
\affiliation{Department of Physics, Faculty of Science, Kyoto University, Kyoto, Kyoto 606-8502, Japan}																									
\author{Grant~P.~Teply}
\affiliation{Department of Physics, University of California, San Diego, La Jolla, CA 92093, USA}																									
\author{Kyohei~Yamada~\orcid{0000-0003-0221-2130}}
\email[Corresponding author: \href{mailto: ykyohei@cmb.phys.s.u-tokyo.ac.jp}{ykyohei@cmb.phys.s.u-tokyo.ac.jp}]{}
\thanks{Current address: Joseph Henry Laboratories of Physics, Jadwin Hall, Princeton University, Princeton, NJ 08544, USA}
\affiliation{Department of Physics, Graduate School of Science, The University of Tokyo, Tokyo 113-0033, Japan}

\collaboration{The \textsc{Polarbear} collaboration}

\date{\today}

\begin{abstract}
The Crab Nebula, also known as Tau A, is a polarized astronomical source at millimeter wavelengths. It has been used as a stable light source for polarization angle calibration in millimeter-wave astronomy.
However, it is known that its intensity and polarization vary as a function of time at a variety of wavelengths. Thus, it is of interest to verify the stability of the millimeter-wave polarization. 
If detected, polarization variability may be used to better understand the dynamics of Tau~A, and for understanding the validity of Tau~A as a calibrator. 
One intriguing application of such observation is to use it for the search of axion-like particles (ALPs). 
Ultralight ALPs couple to photons through a Chern-Simons term, and induce a temporal oscillation in the polarization angle of linearly polarized sources. 
After assessing a number of systematic errors and testing for internal consistency, we evaluate the variability of the polarization angle of the Crab Nebula using 2015 and 2016 observations with the 150\,GHz \textsc{Polarbear} instrument. 
We place a median 95\% upper bound of polarization oscillation amplitude $A < 0.065^\circ$ over the oscillation frequencies from \SI{0.75}{year^{-1}} to \SI{0.66}{hour^{-1}}.
Assuming that no sources other than ALP are causing Tau~A's polarization angle variation, that the ALP constitutes all the dark matter, and that the ALP field is a stochastic Gaussian field, this bound translates into a median 95\% upper bound of ALP-photon coupling $g_{a\gamma\gamma} < 2.16\times10^{-12}\,\mathrm{GeV}^{-1}\times(m_a/10^{-21} \mathrm{eV})$ in the mass range from \SI{9.9e-23}{eV} to \SI{7.7e-19}{eV}. 
This demonstrates that this type of analysis using bright polarized sources is as competitive as those using the polarization of cosmic microwave background in constraining ALPs. 
\end{abstract}

\maketitle

\section{Introduction} \label{sec:intro}
Tau~A\footnote{Also known as M1, NGC 1952, Crab Nebula, and Taurus~A.} is a polarized astronomical source at millimeter wavelengths. It has been used as a stable light source for polarization angle calibration in millimeter-wave astronomy \citep{Aumont:2009dx, Aumont2020, Kam2022}. 
It is also known that its intensity varies as a function of time at a wide variety of timescales and wavelengths \cite{Becker:1995bm, Ellingson:2013mta, Eftekhari:2016fyo, Kouzu:2013mga, MAGIC:2014izm, Feng:2020miz, Bucciantini2023}. 
The precise time-resolved polarimetry of Tau~A is also becoming an active research topic in astrophysics \citep{Moran:2013cla, Feng:2020miz, Bucciantini2023}. 
Thus, it is of interest to verify the stability of the polarization angle of Tau~A in millimeter wavelengths. 
If detected, variability must be considered as the systematic error in the use of Tau~A as a polarization angle calibrator, and may also be important for understanding the dynamics of Tau~A. 

One intriguing application of the time-resolved polarimetry of Tau~A is to use it to search for axion-like particles (ALPs). 
Axions or ALPs, which are pseudo-scalar fields coupled to photons through a Chern-Simons term, are potential candidates for the dark matter. 
The axion was first proposed as a pseudo-Nambu-Goldstone boson arising from a solution to the strong CP problem in quantum chromodynamics \citep{PecceiQuinn1997, PhysRevD.16.1791, Weinberg1978, PhysRevLett.40.279}, while similar particles called ALPs are predicted by string theories \citep{Witten:1984dg, PhysRevD.81.123530, Cicoli:2012sz}.
Although ALPs do not solve the strong CP problems, the ALPs can be bosonic ultralight dark matter and fuzzy dark matter, because their coupling to photons does not have a fixed relationship with their mass~\citep{Ferreira:2020fam, Hui2021}. 
Ultralight dark matter with masses around $10^{-22}$~eV is especially interesting as it could also resolve small scale tensions that exist in the standard cold dark matter model~\citep{FDM2000, Hui:2016ltb}.
Beyond their gravitational effects, ALPs introduce birefringence which presents a unique detection pathway for the presence of ALPs~\citep{Carroll1990, Harari1990}. 
If ALPs are present, one will observe a net rotation of the polarization angle based on the change in the ALP field between where a photon was emitted and detected:
\begin{align}
    \psi = \frac{g_{a\gamma\gamma}}{2}\left(\phi_\mathrm{detected} - \phi_\mathrm{emitted}\right),
    \label{eq:ALP_rotation}
\end{align}
where $g_{a\gamma\gamma}$ is the ALP-photon coupling constant and $\phi$ is the ALP field. The search for polarization angle oscillation due to ALPs has been conducted using the light from astrophysical objects, radio galaxies \citep{Alighieri_2010}, jets in active galaxies \citep{Ivanov:2018byi}, protoplanetary disks \citep{Fujita2019}, pulsars \citep{Liu2020, Basu2021, Castillo_2022, Liu2023}, and the cosmic microwave background (CMB) \citep{Fedderke2019, ALP_BK12, ALP_BK14, ALP_SPT3g} and \citep[][hereafter \citetalias{ALP_PB}]{ALP_PB}.

In this paper, we report our analysis using the observations by the \textsc{Polarbear} experiment, a ground-based experiment installed on the \SI{2.5}{m} aperture oﬀ-axis Gregorian-Dragone type Huan Tran telescope observing the polarization of the CMB from the Atacama Desert in Chile \citep{Kermish:2012eh, Arnold:2010jaq}, at \SI{150}{GHz} from 2015 to 2016. 
This paper is organized as follows.
Section~\ref{sec:ALP} gives the brief review of the phenomenology of the ultralight dark matter and how the polarization oscillation signal appears for the case of Tau~A.
Section~\ref{sec:analysis} describes the procedure of data analysis to calibrate and evaluate polarization angle of Tau~A and its variation.
Section~\ref{sec:validation} describes the data validation methods.
Section~\ref{sec:results} describes the results. 
Section~\ref{sec:systematics} describes dedicated studies for the evaluation of systematic errors.
Discussion and conclusion are described in Sec.~\ref{sec:discussion} and Sec.~\ref{sec:conclusion} respectively.

\section{Modeling of ALP-induced signal} 
\label{sec:ALP}
This section discusses the statistics of ultralight dark matter in general and its phenomenology in the case of ultralight ALPs through the weak interaction with the photons, and models the expected ALP-induced polarization oscillation signal of Tau~A. 
We model the ALP-induced polarization signal assuming that the ALP with a single mass constitutes all the dark matter. 

\subsection{Statistics of the ultralight dark matter}
During the formation of the Milky Way galaxy, the dark matter relaxes into gravitational potential wells and forms dark matter halos with a Maxwellian velocity distribution of characteristic dispersion velocity, $v_\mathrm{vir}{\sim}10^{-3}c$, described by the standard halo model (SHM) \citep{Sikivie1992, PhysRevD.88.123517}. 

We consider the statistical properties of such an ultralight dark matter field, called a virialized ultralight field (VULF) \citep{Derevianko:2016vpm, Foster:2017hbq, Centers:2019dyn}. 
Neglecting the velocity of the dark matter field, the ultralight dark matter field oscillates at its Compton frequency ($=m_ac^2\hbar^{-1}$), where $m_a$ is the mass of dark matter.
The velocity distribution of the SHM leads to the spectral broadening of the oscillation frequency characterized by the coherence time $\tau_\mathrm{coh}$ ($=m_av_\mathrm{vir}^2\hbar^{-1}$), and also leads to the spatial dispersion characterized by the coherence length $\lambda_\mathrm{coh}$ ($=\hbar m_a^{-1} v_\mathrm{vir}^{-1}$).
In the limit that the observation time is much shorter than the coherence time of the dark matter, the time variation of the dark matter field is described as a single mode sinusoid as 
\begin{align}
    \phi(t) = \phi_c \cos\left(\frac{m_ac^2}{\hbar}t\right) + \phi_s\sin\left(\frac{m_ac^2}{\hbar}t\right),
    \label{eq:ALP_field}
\end{align}
where $\phi_c$ and $\phi_s$ are real-valued amplitudes. 
Since the ﬁeld modes of different frequencies and the random phase governed by SHM interfere with each other, the net amplitudes ($\phi_c,~\phi_s$) exhibit stochastic behavior, and follow a Gaussian distribution with variance $\braket{|\phi_c|^2} = \braket{|\phi_s|^2} = \phi_\mathrm{DM}^2/2$. 
The equivalent equation is obtained by the following parametrization:
\begin{align}
    \phi(t) = \phi_0 \sin\left(\frac{m_ac^2}{\hbar}t + \theta \right),
\end{align}
where $\phi_0 = \sqrt{\phi_c^2 + \phi_s^2}$ follows a Rayleigh distribution with scale factor of $\phi_\mathrm{DM}/\sqrt{2}$, and $\theta = \arctan(\phi_s/\phi_c)$ follows uniform distribution from 0 to $2\pi$. 
The $\phi_\mathrm{DM}$ is determined by the average local dark matter density $\rho_\mathrm{DM}{\sim}\SI{0.3}{GeV/cm^3}$ \citep{bovy2012local}, and thus 
\begin{align}
    \phi_\mathrm{DM} = \frac{\hbar}{m_a c}\sqrt{2\rho_\mathrm{DM}}. \label{eq:phiDM}
\end{align}

\subsection{Axion-like polarization oscillation of Tau~A}
\begin{figure}
    \centering
    \includegraphics[width = .45\textwidth]{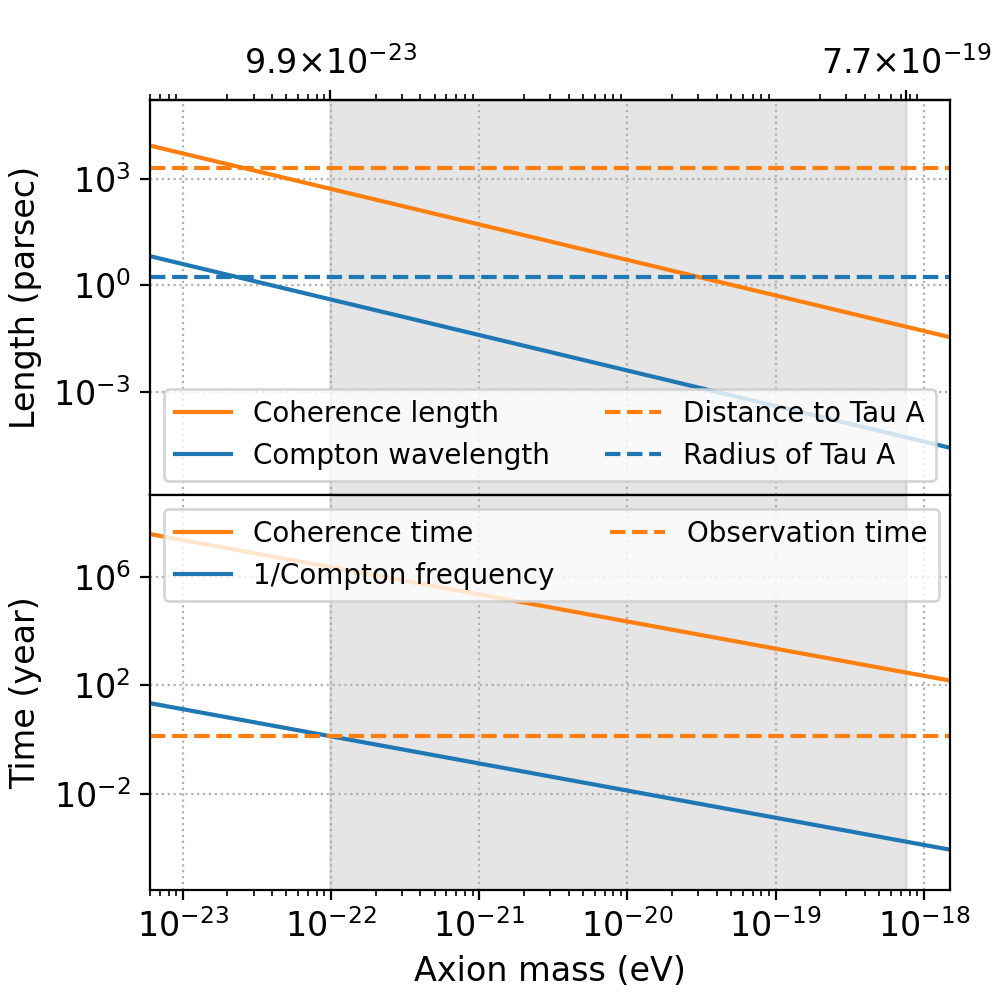}
    \caption{Top: Comparison of typical scales of ALP field and the typical scales of Tau~A. Bottom: Comparison of the typical time scales of ALP field and the observation time of this study. The gray shaded region represents the ALP mass scale of interest in this study.}
    \label{fig:DM_correlation}
\end{figure}
We discuss an axion-like polarization oscillation signal for the case of Tau~A. Tau~A is $2000\pm500$\,pc away from the Earth \citep{Kaplan_2008}, and its diameter is about \SI{1.7}{pc} \citep{Hester2008}.
As shown in Fig.~\ref{fig:DM_correlation}, the distance to Tau~A is much longer than the coherence length of an ALP across the mass scales from \SI{9.9e-23}{eV} to \SI{7.7e-19}{eV}, which is investigated in this study. Thus the oscillation amplitudes of the local field and the field at Tau~A are independent.
Since the size of Tau~A is much larger than the Compton wavelength ($=\hbar m_a^{-1} c^{-1}$) of an ALP, the oscillation at Tau~A averages out, because various phases of oscillation at Tau~A contributes to the signal \citep[See Eq.~(2.18) of][]{Chigusa2020}.  
Therefore, the search of an axion-like polarization oscillation by Tau~A is only sensitive to the local ALP field, and the signal (Eq.~\eqref{eq:ALP_rotation}) may be modeled as 
\begin{align}
    \psi(t) = \psi_0 + \frac{g_{a\gamma\gamma}}{2}\phi_\mathrm{local}(t),
\end{align}
where $\psi_0$ is the polarization angle of Tau~A, and $\phi_\mathrm{local}(t)$ is the local stochastic ALP field, which oscillates according to Eq.~\eqref{eq:ALP_field}.

\section{Analysis}\label{sec:analysis}
The data processing and map-making pipeline follows \citet[][hereafter \citetalias{PB20}]{PB20} with the improved of half-wave plate (HWP) angle estimation introduced in the improved result \cite{PB22}.
The data set of this study was analyzed for the calibration of \citetalias{PB20}.
In this study, we re-analyze the same data set adopting a blind-analysis framework \citep{Klein:2005di}.
The polarization angle of Tau~A in each observation is not seen (blinded) until the validation of the data analysis is completed (Sec.~\ref{sec:validation}), i.e. the criteria for the internal consistency test (Sec.~\ref{sec:nulltest_result}) are met, and the possible systematic errors are evaluated (Secs.~\ref{sec:syst_correlation} and \ref{sec:systematics}). 
The polarization maps and the absolute angle of Tau~A ($\psi_0$) averaged over the entire observation period are not blinded. 

\subsection{Observations} \label{sec:observations}
\begin{figure}
    \centering
    \includegraphics[width = .45\textwidth]{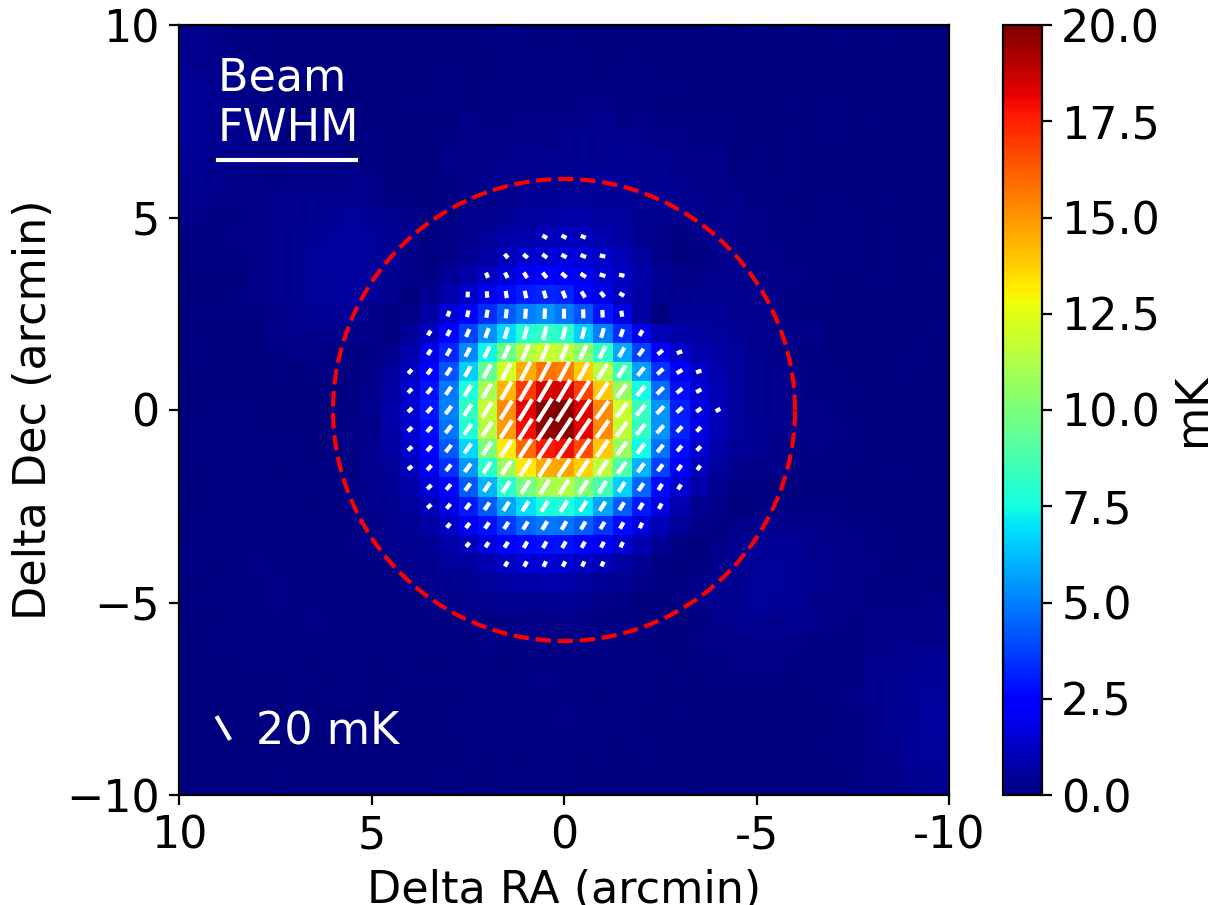}
    \caption{Polarization map of Tau~A observed by \textsc{Polarbear} in 4th and 5th season from 2015 to 2016. The color scale represents the polarization intensity of Tau~A, $\sqrt{Q_\mathrm{Tau A}^2+U_\mathrm{Tau A}^2}$. The orientations of white bars in map pixels represent the average polarization angles ($\psi_0$) at each map pixel. The red dashed circle represents the integration region to estimate average Tau~A polarization angle or to evaluate systematic errors, which is consistently used in this study.}
    \label{fig:taua_map}
\end{figure}
In the 4th and 5th seasons, over 486 days from 2015 to 2016, \textsc{Polarbear} observed Tau~A several times per week, for a total of 220 observations. 
The \textsc{Polarbear} receiver employs a total of 1274 transition-edge sensor (TES) bolometers cooled to 0.3 K observing the sky through lenslet-coupled dipole antennas \citep{Arnold:2012sr}.
While \textsc{Polarbear} has a beam with a 3.6$^\prime$ FWHM, Tau~A has an angular size of $7^\prime\times5^\prime$~\citep{GreenSNRcatalogue2019}.
Therefore, we do not examine the detailed structure of Tau~A.
Figure \ref{fig:taua_map} shows the map of Tau~A observed by \textsc{Polarbear}. Throughout this study, the polarization angle of Tau~A is evaluated by integrating the map around Tau~A within 12$^\prime$ diameter (Sec. \ref{sec:map2angle}).

Tau~A is at declination of 22$^\circ$ and \textsc{Polarbear} is at $-23^\circ$ latitude. Therefore, the highest elevation of Tau~A observed by \textsc{Polarbear} is 45$^\circ$. 
In our observing schedule, \textsc{Polarbear} always observes Tau~A when it sets, from 39$^\circ$ to 30$^\circ$ in elevation, using a same raster scan pattern. 
The telescope scans 5.5$^\circ$ back-and-forth in azimuth at 0.2$^\circ$/s, while the elevation continuously tracks Tau~A. 
Between each sweep in azimuth, called a subscan, the elevation changes by 2$^\prime$, about half of the beam width. 
The typical observation time for each observation is one hour.

Before and after every observation, a chopped thermal source calibration is carried out to characterize the relative detector gain, gain drift, and detector time constant. 
These values can be different for each observation depending on how each TES was tuned which depends on optical conditions of the atmosphere. 

\subsection{Data selection}
\label{sec:dataselection}
We apply the same data selection criteria as \citetalias{PB20} except for the following optimization of the data selection conditions. 
The data volume is summarized in Table~\ref{table:data_volume}. 
Observation efficiency is low because the primary science target of \textsc{Polarbear} is the CMB, not Tau~A. 
Since \textsc{Polarbear} has a 2.4$^\circ$ field of view, the Tau~A signal is contained in less than 1\% of the entire data. 
The remaining data outside of 12$^\prime$ diameter around Tau~A are used for evaluating the data quality. 
The data selection efficiency is summarized in Table~\ref{table:data_efficiency}. 
After the data selection, the number of observations is reduced from 220 to 95.

The data selection criteria for Tau~A observations differ from \citetalias{PB20} in three ways.
First, the threshold for weather conditions is tightened 
to average precipitable water vapor (PWV) $<$ \SI{2.5}{mm}, and newly introduced criterion allowing data only when max PWV $-$ min PWV $<$ \SI{0.6}{mm}. These criteria are motivated by the estimated intensity to polarization leakage (Sec. \ref{sec:cal_I2P}).
Second, the threshold for common mode power spectral density (PSD) is relaxed, because the common mode atmospheric signal is larger in Tau~A observations than in CMB observations.
During the observation, the telescope tracks Tau~A and changes elevation, CMB observations are performed with the constant elevation scans (CES).
Third, the Sun-Moon avoidance cut is relaxed 
to allow Tau~A observations more than $20^\circ$ away from either source, because the observation patch for Tau~A is smaller than the CMB patch in \citetalias{PB20}.

\begin{table*}
    \centering
    \begin{minipage}{0.45\linewidth}
    \caption{Data volume.}
    \label{table:data_volume}
        \begin{tabular*}{\textwidth}{@{\extracolsep{\fill}}lc}
        \hline \hline
        Observation & \\
        ~~from & 2015 Aug 25 \\
        ~~until & 2016 Dec 20 \\
        Total calendar time & \SI{11616}{hr} \\
        Time observing Tau~A & \SI{243}{hr} \\
        Observation efficiency & 2.1\% \\
        \hline
        Total number of detectors & 1274 \\
        Calibrated detectors & 647 \\
        Detector yield & 50.8\% \\
        \hline
        Total volume of data & \SI{156,998}{hr} \\ 
        Final volume of data & \SI{47,210}{hr} \\
        Data selection efficiency & 30.1\%\\
        \hline
        Overall efficiency & 0.63\% \\
        \hline
        \hline
    \end{tabular*}
    \vspace{0ex}
    \end{minipage}
    \begin{minipage}{0.1\linewidth}
    \end{minipage}
    \begin{minipage}{0.45\linewidth}
    \centering
    \caption{Data selection efficiency.}
     \label{table:data_efficiency}
        \begin{tabular*}{\textwidth}{@{\extracolsep{\fill}}ll}
        \hline \hline
        Stage of data selection & \\
        \hline
        \textbf{Selection of observations} & \\
        Terminated observation & 98.1\% \\
        Detector stage temperature & 92.9\% \\
        Weather condition & 84.5\% \\
        Sun Moon distance & 78.7\% \\
        Instrumental problem & 88.4\% \\ \hline
        \textbf{Selection within one observation} & \\
        Non-operating detectors & 79.4\% \\
        Packet drop & 98.3\% \\
        Individual detector glitch & 99.6\% \\
        Common mode glitch & 89.8\% \\
        Individual detector PSD & 85.9\% \\
        Common mode PSD & 94.5\% \\
        \hline
        Cumulative data selection & 30.1\% \\
        \hline
        \hline
    \end{tabular*}
    \end{minipage}
\end{table*}

\subsection{Time domain processing and demodulation}
\label{sec:TODprocessing}
This section provides a brief review of the processing of the timestreams, measured by the TESs in the \textsc{Polarbear} receiver. Each TES bolometer is coupled to a dipole-slot antenna and measures a single polarization of incident light.
\textsc{Polarbear} employs a half-wave plate (HWP) at the prime focus which continuously rotates at \SI{2}{Hz}. 
Thus, the idealistic timesteams of the detector are modulated as follows \citep[][hereafter \citetalias{Takakura:2017ddx}]{Takakura:2017ddx}:
\begin{align}
    d_m(t) = I(t) + \mathrm{Re}[(Q(t) + iU(t))m(\chi)],
\end{align}
where $I(t)$, $Q(t)$ and $U(t)$ are the incident Stokes parameters defined in telescope coordinate, $\chi(t)$ is the rotation angle of the HWP, and
\begin{align}
    m(\chi) = \exp(-i4\chi)
\end{align}
is the modulation function.
We obtain the intensity signal $d_0(t)=I(t)$ by applying a low-pass filter, and also we demodulate the timestreams and extract the polarization signal as $d_d^\mathrm{ideal}(t) = Q(t) + iU(t)$.
In the presence of the static instrumental polarization ($A_4$), the mis-estimation of the HWP angle ($\Delta\chi$), the polarization angle of detector ($\theta_\mathrm{det}$), the intensity to polarization (I2P) leakage coefficient ($\lambda_4$), the time constant of detector ($\tau$), and the timing offset ($\Delta t$) the modulated timestreams is modified as
\begin{align}
    &d_m(t) = I(t) + \mathrm{Re}[(A_4 + \lambda_4\Delta I(t) + d_d^\mathrm{ideal})m'(\chi)],
    \label{eq:modified_mod}
\end{align}
where $\Delta I(t) = I(t) - \braket{I(t)}_t$ is the drift of the intensity signal, and the modified modulation function $m'(\chi)$ is
\begin{align}
    m'(\chi) &= \exp \left(-i4\chi(t-\tau-\Delta t) + i4\Delta\chi + i2\theta_\mathrm{det}\right) \nonumber \\
    &\simeq \exp \left(-i4\chi(t) + i4\dot{\chi}\tau + i4\dot{\chi}\Delta t + i4\Delta\chi + i2\theta_\mathrm{det}\right).
    \label{eq:modified_modfunction}
\end{align}
Therefore, the demodulated timestreams are modified as
\begin{align}
    &d_d(t) \simeq (A_4 + \lambda_4\Delta I(t) + d_d^\mathrm{ideal})e^{i4\dot{\chi}\tau+i4\dot{\chi}\Delta t+i4\Delta\chi+i2\theta_\mathrm{det}}. 
\end{align}
Note that $A_4$ and $\lambda_4$ are complex values, and others are real values.
$A_4$, $\Delta\chi$ and $\Delta t$ are common to all detectors, while $\theta_\mathrm{det}$, $\lambda_4$ and $\tau$ are different. 
Also, $\arg(A_4)$ and $\theta_\mathrm{det}$ are assumed to be stable throughout all the observations, while others may vary observation-by-observation.
The effects of time constant $\tau$ and polarization angle of detector $\theta_\mathrm{det}$ are deconvolved, and I2P leakage effect is subtracted, in time domain. This leaves 
\begin{align}
    \begin{split}
    d_d(t) \simeq (A_4 + \Delta\lambda_4\Delta I(t) + d_d^\mathrm{ideal}) \\
    \quad \times e^{i4\dot{\chi}\Delta\tau+i4\dot{\chi}\Delta t+i4\Delta\chi+i2\Delta\theta_\mathrm{det}},
    \end{split}
    \label{eq:modified_demod}
\end{align}
where $\Delta \tau$, $\Delta \theta_\mathrm{det}$ and $\Delta \lambda_4$ are the mis-estimation of $\tau$, $\theta_\mathrm{det}$ and $\lambda_4$, respectively.
The argument of complex phase of Eq.~\eqref{eq:modified_demod} corresponds to the mis-estimation of polarization angle in telescope coordinate, therefore the corresponding mis-estimation of the polarization angle of Tau~A is expressed as 
\begin{align}
    \Delta\psi &= 2\left(\Delta \chi + \dot{\chi}\Delta t + \dot{\chi}\Delta \tau\right) + \Delta\theta_\mathrm{det}.
    \label{eq:angle_bias}
\end{align}
Also, the mis-estimation of the I2P leakage coefficient $\Delta\lambda_4$ biases the Tau~A polarization angle as
\begin{align}
    \psi + \Delta\psi &= \frac{1}{2}\tan^{-1} \left( \frac{U_\mathrm{Tau~A} + \Delta\lambda_4^\mathrm{imag} I_\mathrm{Tau~A}}
    {Q_\mathrm{Tau~A} + \Delta\lambda_4^\mathrm{real} I_\mathrm{Tau~A}}\right),
    \label{eq:I2P_bias}
\end{align}
where $Q_\mathrm{Tau~A}$, $U_\mathrm{Tau~A}$ and $I_\mathrm{Tau~A}$ are the Stokes parameters of the incident light from Tau~A. This bias is approximately expressed as
\begin{align}
    |\Delta\psi| \lesssim \frac{1}{2}\frac{|\Delta\lambda_4|}{|p_\mathrm{frac}|},
    \label{eq:I2P_to_syst}
\end{align}
where $p_\mathrm{frac}$ is a complex polarization fraction of Tau~A defined as $p_\mathrm{frac}I_\mathrm{Tau~A} = Q_\mathrm{Tau~A} + iU_\mathrm{Tau~A}$.
Each of these calibrations and imperfections are characterized as we describe in the following section.

\subsection{Calibration}
\label{sec:calibration}
This section highlights the improved polarization angle calibration methods in this study.
The calibration of the pointing model of the telescope, pointing offsets of detectors, effective beam function, relative gain variations, detector time constants, and polarization efﬁciencies are the same as those used in \citetalias{PB20}.

\subsubsection{Relative polarization angle between detectors}
\label{sec:polcaldet}
The relative polarization angle between detectors is calibrated from the average of Tau~A polarization angle ($\psi_0$) over the entire observation period for each detector in the same way as \citetalias{PB20}, with all the improvements of the I2P evaluation and the calibration of relative polarization angle between observations described in the following sections. 
The calibration of relative polarization angle between detectors is performed iteratively until it converges. 
Typically two iterations are required. 
The uncertainty of the relative polarization angle between detectors is estimated to be 0.1$^\circ$ for each detector, which is negligibly small as we discuss in Sec \ref{sec:syst_detangle}.

\subsubsection{Intensity to polarization leakage}
\label{sec:cal_I2P}
The estimation of the I2P leakage is updated from \citetalias{PB20}. 
To validate the I2P leakage estimates, two different I2P leakage estimates are performed on Jupiter observations and the results are compared. There are two major sources of I2P leakage, one is a leakage due to the imperfection of optical components and the other is due to the nonlinearity of the detectors (\citetalias{Takakura:2017ddx}). 
The I2P leakage estimated by two different methods includes both effects.

The first method uses the unpolarized atomospheric $1/f$ signal, which closely follows the method used in \citetalias{PB20} developed by \citetalias{Takakura:2017ddx}. 
We take the correlation between the one-hour intensity timestream and one-hour polarization timestream and obtain the I2P leakage coefficient $\lambda_4$.
This assumes that the intensity timestream is dominated by the signal of unpolarized atmospheric fluctuation, and the fluctuation of the polarization timestream is dominated by the intensity timestream leaked into polarization by the I2P leakage effect. 
Tau~A, or Jupiter for validation, is masked out from the timestream so that the Tau~A (Jupiter) signal does not bias the estimation of $\lambda_4$. 

The second method uses Jupiter signal.
The I2P can be decomposed into monopole, dipole, and quadrupole terms \citep{ABS:2016rpo}.
We measure the monopole term by integrating the map around Jupiter within 12$^\prime$ diameter as
\begin{align}
    \lambda_4^\mathrm{real} &= \frac{\sum_p^\mathrm{\diameter < 12^\prime}Q_p^\mathrm{Jupiter}}{\sum_p^\mathrm{\diameter < 12^\prime}I_p^\mathrm{Jupiter}} \\
    \lambda_4^\mathrm{imag} &= \frac{\sum_p^\mathrm{\diameter < 12^\prime}U_p^\mathrm{Jupiter}}{\sum_p^\mathrm{\diameter < 12^\prime}I_p^\mathrm{Jupiter}},
\end{align}
where $I_p$, $Q_p$ and $U_p$ are the $p$-th I, $Q$ and $U$ map pixel.
The contamination from higher order multipoles is found to be small \citepalias{Takakura:2017ddx}.
Here we assume that the polarization of Jupiter is purely from I2P from our instruments. 
Jupiter is slightly polarized due to synchrotron radiation \citep{Pater1981, VLA2003}, and its polarization at \SI{150}{GHz} is expected to be $\ll 1\%$.

\begin{figure}
    \centering
    \includegraphics[width = .45\textwidth]{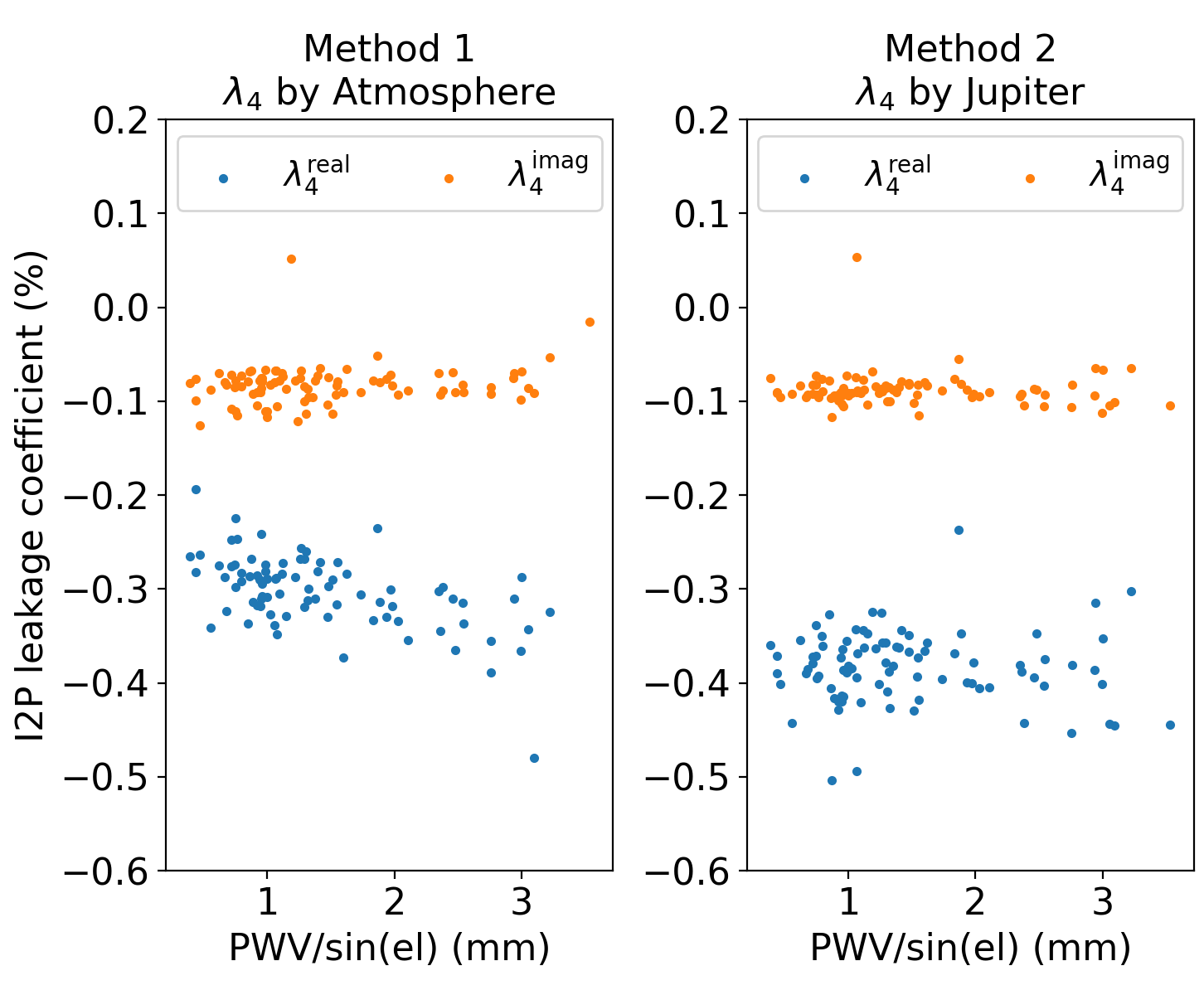}
    \caption{Comparison of the PWV dependence of the two intensity to polarization coefficient estimates $\lambda_4$ for Jupiter observations. Each point is a median of all detectors for one observation. The PWV is scaled by sin(el) to compare the line of sight integrated value.}
    \label{fig:I2P_pwv}
\end{figure}
Figure \ref{fig:I2P_pwv} shows the comparison of the observation-by-observation $\lambda_4$ by the two methods.
The $\lambda_4$ of method~1 shows larger PWV dependence than that of method~2, and the $\lambda_4$ at larger PWV show closer values to that of method~2.
We attribute the PWV dependence of method~1 to bias, based on the verification of the systematic errors discussed in Sec~\ref{sec:syst_correlation}.

The I2P estimates by method~1 may be biased, because at lower PWV, the $1/f$ fluctuation from non-optical effects, such as the fluctuation of the focal plane temperature or the temperature of the primary or secondary mirror, the warm readout electronics may not be negligible \citep{Dtanabe2022}\footnote{ The non-optical $1/f$ fluctuation can be larger in the Tau~A or Jupiter scans than in the CES scans of \citetalias{PB20}, because it tracks the sources by changing elevation. }.
On the other hand, method 2 is robust to any $1/f$ fluctuations because it is estimated by the point source, Jupiter.
However, method 2 also has its drawbacks, such as the fact that they are observed under different observing conditions and that Jupiter is about 10 times brighter ($T_\mathrm{Jupiter}\sim$\SI{2}{K}) than Tau~A, so it may be more affected by the detector non-linearities and have constant bias.

\begin{figure}
    \centering
    \includegraphics[width = .45\textwidth]{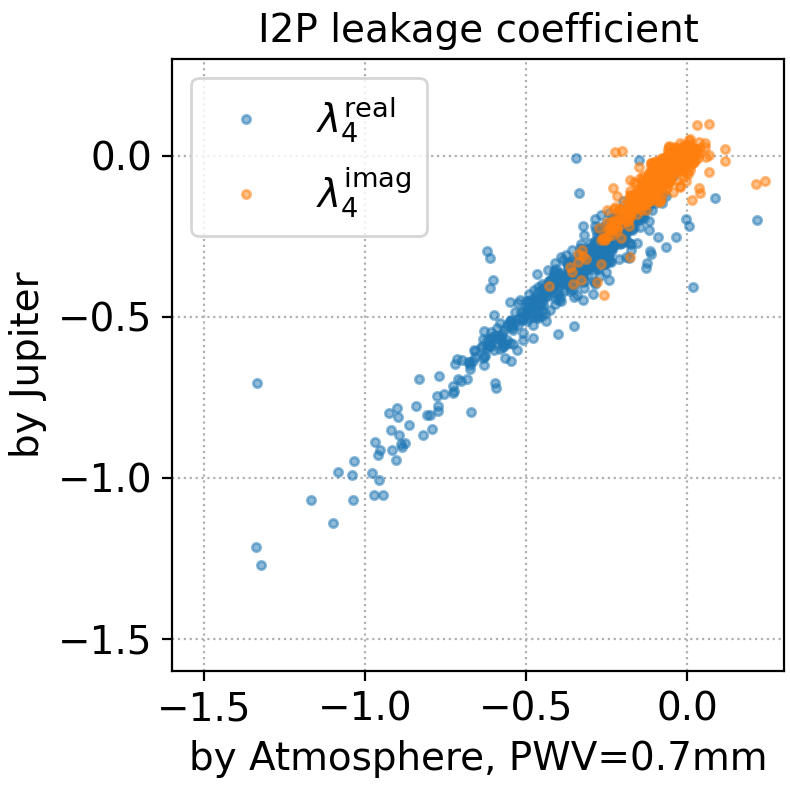}
    \caption{Comparison of the each detector's intensity to polarization coefficient $\lambda_4$ estimated by the Jupiter signal and the atmospheric $1/f$ signal in Tau~A observations.}
    \label{fig:I2P_det}
\end{figure}

In our Tau~A analysis, we construct the I2P leakage model by making a modification to method~1 and applying it to the Tau~A map-making process. We model the PWV dependence of method~1 as a linear function for each detector and substitute the median PWV (=\SI{0.7}{mm}) of Tau~A observations, to make representative $\lambda_4$ for each detector. 
We call this method~1$^\prime$.
Figure \ref{fig:I2P_det} shows the comparison of detector-by-detector I2P coefficient $\lambda_4$ by method~1$^\prime$ and method~2. They show reasonable agreement. 
We discuss the uncertainty of I2P leakage estimates by the difference of method 1$^\prime$ and 2 in Sec.~\ref{sec:systematics}.

\subsubsection{Relative polarization angle between observations}
\label{sec:polcalobs}
As we have shown in Eq.~\eqref{eq:angle_bias}, a mis-calibration of the relative polarization angle between observations may be caused by the mis-estimation of the HWP angle ($\Delta\chi$), the time constant of detectors ($\Delta \tau$), and the timing offset ($\Delta t$).
In \citetalias{PB20}, we calibrated the time constant of each detector in each observation by the average of the chopped thermal source calibration right before and after the Tau~A observations, and the achieved polarization angle error was sufficiently small. 

In the search for polarization oscillation, the importance of the calibration of relative polarization angle between observations increases compared to the analysis which takes an average of all observations, e.g. CMB power spectrum analysis in \citetalias{PB20}.
In this study, we further calibrate the relative polarization angle between observations assuming that the telescope and the polarization angle of the instrumental polarization due to the primary mirror ($\frac{1}{2}\arg(A_4)$) are stable.\footnote{The instrumental polarization is not modulated by the ALP, because the distance between the primary mirror and the detectors ($\sim$1~m) is much shorter than the coherence length of local ALP field in the mass scale of our interests (${\sim}10^{15}-10^{18}$\,m) }
Because our HWP modulator is the second optical element in the path of light from the sky to the focal plane \citepalias{Takakura:2017ddx}, the reflection at the primary mirror produces the instrumental polarization \citep{PIQUE_CAPMAP_2005}.
This becomes a good calibrator for the relative polarization angle between observations, because it illuminates all detectors stably and uniformly ($|A_4|{\sim}$\SI{0.1}{K}, \citetalias{Takakura:2017ddx}).
The instrumental polarization is estimated as a signal synchronous to the 4th harmonic of the rotation angle of the HWP, by averaging the modulated timestreams (Eq.~\eqref{eq:modified_mod}) over the HWP angle for each observation \citepalias{Takakura:2017ddx}, masking Tau~A signal and glitches in the time domain. 

The amount of the corrected polarization angle variation between observations is $0.16^\circ$ root mean square (RMS). 
This variation is measured within the statistical uncertainty of the instrumental polarization angle $0.02^\circ$, which is a median of the statistical uncertainty per observation in terms of the standard deviation (STD).



\subsection{Estimation of polarization angle of Tau~A and its statistical uncertainty} \label{sec:map2angle}
The polarization angle of Tau~A ($\psi$) is estimated in the map-domain for each observation, with the same filtered-binned map-making as \citetalias{PB20}. 
We estimate $\psi$ by integrating over the map around Tau~A within 12$^\prime$ diameter as  
\begin{align}
    d_t = \frac{1}{2}\tan^{-1} \left( \frac{\sum_p^{\diameter<12'} U_p^\mathrm{Tau~A}}{\sum_p^{\diameter<12'} Q_p^\mathrm{Tau~A}}\right),
    \label{eq:angle_from_map}
\end{align}
where $d_t$ is the measured $\psi$, and $t$ is the average observation time over one hour.
The diameter of the integration region is chosen so that the region covers the size of Tau~A convolved by our beam size (FHWM~3.6$^\prime$) including the measured boresight pointing drift of our telescope (Sec.~\ref{sec:syst_pointing}). This angle estimation method is not the most efficient method, but it is immune to the complexity of the structure of Tau~A, and robust to systematic errors.

The statistical uncertainty of each observation $\sigma_t$ was estimated by the bootstrap method, also called ``signflip noise estimation," which is widely used for noise estimation in the CMB analysis. 
First, we generate the signflip noise only map for each observation by randomly assigning a $+1$ or $-1$ factor to each detector and co-adding individual detector maps\footnote{The same random sign patterns are used for all the null test data-splits (Sec.~\ref{sec:validation}) to ensure the correlations between the data-splits are accounted correctly.}. 
The random sign is assigned so that the total data weight is even
for both signs. The signflip map will cancel the Tau~A signal as well as the $1/f$ noise which is common among all the detectors. 
Then, we sample the noise realizations of Stokes parameters of $Q$ and $U$ from signflip map over independent 48 regions with the same size as the one used to estimate $\psi$ as $\delta Q = \sum_p^{\mathrm{noise~region}}Q_p^\mathrm{Noise}$, $\delta U = \sum_p^{\mathrm{noise~region}}U_p^\mathrm{Noise}$.
One of the statistical noise realizations of Tau~A polarization angle $\delta \psi$ is obtained by adding $\delta Q$ and $\delta U$ to the Tau~A signal as
\begin{align}
    \psi + \delta \psi = \frac{1}{2}\tan^{-1} \left(
    \frac{\sum_p^{\diameter<12'} U_p^\mathrm{Tau~A}+\delta U}
    {\sum_p^{\diameter<12'} Q_p^\mathrm{Tau~A}+\delta Q}
    \right).
    \label{eq:signflip_noise}
\end{align}
The statistical uncertainty $\sigma_t$ is obtained from the STD of the bootstrapped noise realizations $\delta \psi$.
The typical value of $\sigma_t$ is $0.16^\circ$ from the median of all observations.

\subsection{Estimation of Tau~A polarization angle spectrum}\label{sec:spectrum}

As discussed in Sec.~\ref{sec:observations}, the Tau~A observations span over 486 days, and are performed daily.
Therefore, the Nyquist frequency ($f_\mathrm{NYQ}$) is well-defined in our data set.
Because of the aliasing, the frequency bins above $f_\mathrm{NYQ}$ are strongly correlated with the frequency bins below $f_\mathrm{NYQ}$,  
and almost all degrees of freedom in our data are contained in the following set of 242 bins:
\begin{align}
    f_\mathrm{bins}=\left\{\frac{1}{486\Delta T}, \frac{2}{486\Delta T}, ..., \frac{242}{486\Delta T}\right\}
    \label{eq:fbins}
\end{align}
where $\Delta T \simeq 1-1/366.24$\,day is the observation interval, one local sidereal day.
The harmonics of $f_\mathrm{NYQ}(=1/(2\Delta T))$ are not included, because the harmonics of $f_\mathrm{NYQ}$ correspond to the constant mode and we do not have sensitivity to them.
Over the 486~days of the observation period, we have 95 observations of Tau~A. Therefore, the number of frequency bins (242) is larger than the effective number of degrees of freedom (95).
This induces a bin-by-bin correlations among the frequency bins below $f_\mathrm{NYQ}$.
We also accept the bin-by-bin correlations due to aliasing, and the scientific result is extended to 
\begin{align}
    f_\mathrm{bins, result} = \left\{\frac{N}{2\Delta T} + f_\mathrm{bins}  \middle| N = 0, 1, ..., 31\right\}.
    \label{eq:fbins_science}
\end{align}
The maximum frequency to which we have sensitivity is determined by the duration of a single Tau~A observation, one hour (Sec.~\ref{sec:observations}).

We estimate the frequency spectrum by least-square spectral analysis \citep{LSST1969}, similar to Fourier analysis. Hereafter we call the amplitude of least-square spectral analysis ``LSSA."  
The LSSA of the measured Tau~A angle $d_t$ with the error bar $\sigma_t$ is
\begin{align}
    \tilde{d}_f = \underset{A\in\mathcal{C}}{\operatorname{argmax}}\left[\exp\left\{-\frac{1}{2}\sum_t\left(\frac{d_t-\mathrm{Re}(Ae^{-i2\pi f t})}{\sigma_t}\right)^2\right\}\right],
    \label{eq:LSSA}
\end{align}
where $\underset{A\in\mathcal{C}}{\operatorname{argmax}}[f(A)]$ represents a complex amplitude that maximizes a function $f(A)$.
The solution of Eq.~\eqref{eq:LSSA} is
\begin{align}
    \begin{pmatrix}
    \tilde{d}_f^\mathrm{real} \\ \tilde{d}_f^\mathrm{imag}
    \end{pmatrix}
    =
    \begin{pmatrix}
    \sum_{t'} \frac{c_{t'}^2}{\sigma_{t'}^2} & \sum_{t'} \frac{c_{t'}s_{t'}}{\sigma_{t'}^2} \\
    \sum_{t'} \frac{c_{t'}s_{t'}}{\sigma_{t'}^2} & \sum_{t'} \frac{s_{t'}^2}{\sigma_{t'}^2} 
    \end{pmatrix}^{-1}
    \begin{pmatrix}
    \sum_t \frac{c_t d_t}{\sigma_t^2} \\ \sum_t \frac{s_t d_t}{\sigma_t^2}
    \end{pmatrix},
    \label{eq:LSSA_real_matrix}
\end{align}
where $c_t=\cos(2\pi ft), s_t=\sin(2\pi ft)$.
Hereafter, we express the complex amplitude of the LSSA as 
\begin{align}
    \tilde{d}_f \equiv \tilde{d}_f^\mathrm{real} +i\tilde{d}_f^\mathrm{imag} = \sum_t F_{ft}d_t,
    \label{eq:LSSA_complex_matrix}
\end{align}
where
\begin{align}
    F_{ft} = \begin{pmatrix} 1\\i \end{pmatrix}^\mathrm{T}
    \begin{pmatrix}
    \sum_{t'} \frac{c_{t'}^2}{\sigma_{t'}^2} & \sum_{t'} \frac{c_{t'}s_{t'}}{\sigma_{t'}^2} \\
    \sum_{t'} \frac{c_{t'}s_{t'}}{\sigma_{t'}^2} & \sum_{t'} \frac{s_{t'}^2}{\sigma_{t'}^2} 
    \end{pmatrix}^{-1}
    \begin{pmatrix}
    \frac{c_t}{\sigma_t^2} \\ \frac{s_t}{\sigma_t^2}
    \end{pmatrix}
\end{align}
Similarly, in the following we express the LSSA of the ALP oscillation and noise as $\tilde{\phi}_f = \sum_t F_{ft}\phi_t$ 
and $\tilde{n}_f =  \sum_t F_{ft} n_t$, respectively. Note that $\tilde{\phi}_f$ and $\tilde{n}_f$ are defined using the same noise weight, $F_{ft}$.

The LSSA estimate of the signal amplitude at each frequency is biased for multiple reasons.
We quantify the bias on the oscillation amplitude averaging over the phase of the oscillation by the transfer function $F_f$, the detailed derivation of which is described in Appendix \ref{apx:transfer_function}.
Figure~\ref{fig:transferfunction} represents the individual transfer functions and the overall transfer function. The overall transfer function is evaluated by simulating all the transfer functions at once.
\begin{figure}
    \centering
    \includegraphics[width = 0.5\textwidth]{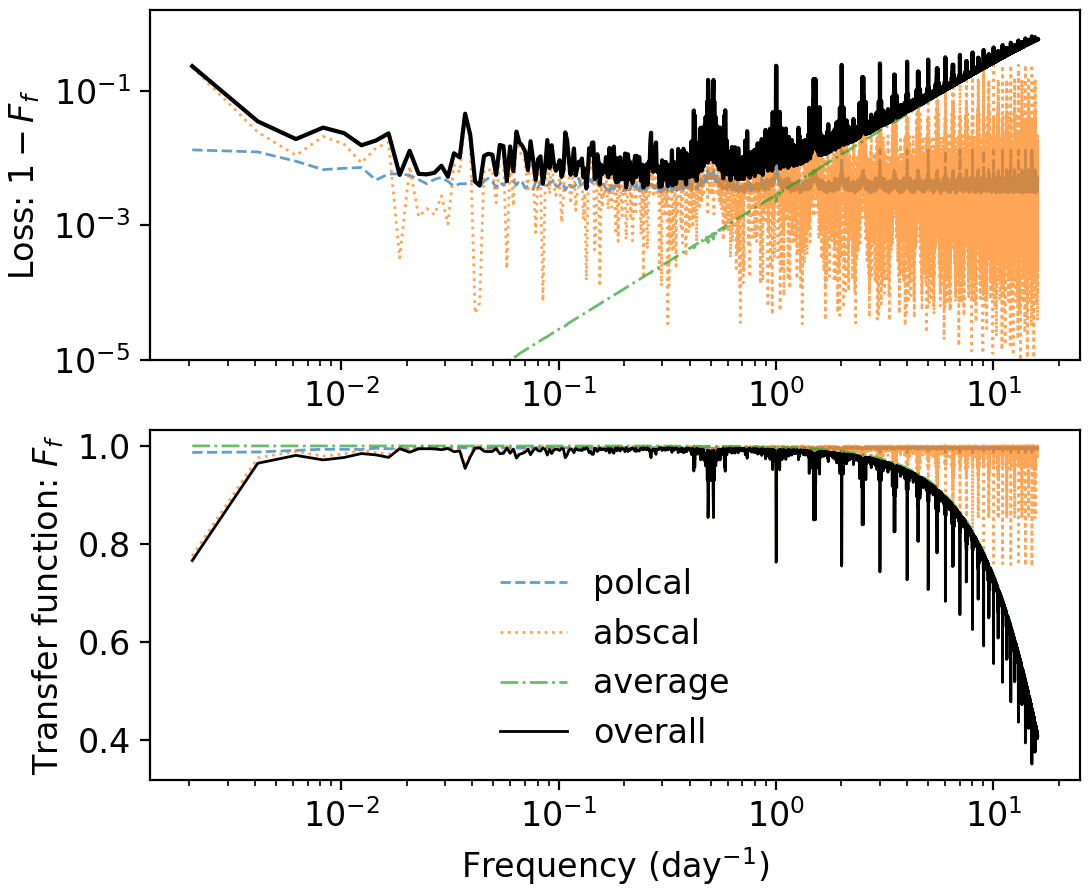}
    \caption{Overall transfer function and individual transfer functions over the frequency bins (Eq.~\eqref{eq:fbins_science}). The transfer function is averaged over all the phases of oscillations. Bottom panel shows the transfer function $F_f$, the top panel shows $1-F_f$. The calibration of relative polarization angle between detectors using Tau~A has the effect of reducing the signal almost uniformly to 99.5\% (polcal).
    The absolute polarization angle ($\psi_0$) calibration has the effect of reducing the signal at certain frequencies that our observing schedule does not favor (abscal). The averaging over one-hour observation time has the effect of reducing the signal at high frequencies (average).}
    \label{fig:transferfunction}
\end{figure}

\subsection{Estimation of polarization oscillation amplitude}
We search for a single mode polarization oscillation without knowing its frequency and phase. Therefore, we take the convolution for all possible phases, whose probability density function is flat, and estimate the amplitude $A_f^\mathrm{obs}$ assuming the presence of a signal at each frequency.
We maximize the following likelihood function: 
\begin{align}
    P(\tilde{d}_f | A_f) = \int_0^{2\pi} d\theta_f~P(\tilde{d}_f|A_f, \theta_f) P(\theta_f),
    \label{eq:A_likelihood}
\end{align}
where $A_f$ and $\theta_f$ the real-valued true amplitude and phase of the oscillation signal at $f$, and $P(\theta_f)$ is the uniform probability density.
\begin{align}
    P(\tilde{d}_f|A_f, \theta_f) = 
    \frac{\exp\left(-\frac{1}{2} \underset{m,n\in\set{\mathrm{real},\mathrm{imag}}}{\sum}
    \tilde{\delta}_f^m\tilde{N}_{f,mn}^{-1}\tilde{\delta}_f^n \right)
    }{\sqrt{(2\pi)^2\,\mathrm{det}(\tilde{N}_f)}}
    \label{eq:Pdf_given_parameters}
\end{align}
is the probability density for obtaining $\tilde{d}_f$ for given signal parameters, $A_f$ and $\theta_f$. 
\begin{align}
    \tilde{\delta}_f \equiv \tilde{d}_f - F_fA_f\exp(i\theta_f), 
    \label{eq:data-signal}
\end{align}
is the difference of the LSSA of data minus signal, and $\tilde{N}_f$ is a noise covariance matrix with elements  
\begin{align}
    \tilde{N}_f = \begin{pmatrix}
     \braket{|\tilde{n}_f^\mathrm{real}|^2} & \braket{\tilde{n}_f^\mathrm{real} \tilde{n}_f^\mathrm{imag}} \\
     \braket{\tilde{n}_f^\mathrm{imag} \tilde{n}_f^\mathrm{real}} & \braket{|\tilde{n}_f^\mathrm{imag}|^2}
    \end{pmatrix},
    \label{eq:ncov}
\end{align}
where $\braket{\cdot}$ represents average over signflip noise simulations.
Note that when the noise covariance is diagonal and $\braket{|d_f^\mathrm{real}|^2} = \braket{|d_f^\mathrm{imag}|^2}$, the Eq.~\eqref{eq:A_likelihood} is simplified as
\begin{align}
    P(\tilde{d}_f|A_f) = 
    \frac{I_0 \left(\frac{2A_f|\tilde{d}_f|}{\braket{|\tilde{n}_f|^2}} \right)}{\pi\braket{|\tilde{n}_f|^2}} 
    \exp\left(-\frac{|\tilde{d}_f|^2+F_f^2A_f^2}{\braket{|\tilde{n}_f|^2}}\right).
\end{align}
where $I_0$ is the modified Bessel function of the first kind. The equivalent likelihood is obtained by considering that $2|\tilde{d}_f|^2/\braket{|\tilde{n}_f|^2}$ follows the non-central chi-square distribution with non-centrality of $2F_f^2A_f^2/\braket{|\tilde{n}_f|^2}$.

To evaluate the detection significance, we form the following test statistic, which quantifies the global significance of the preference for the signal over the null hypothesis:
\begin{align}
    \Delta \chi^2 &= \mathrm{max}_f\left( \Delta \chi^2_f \right),
\end{align}
where
\begin{align}
    \Delta \chi^2_f &= -2\log\left( \frac{P(\tilde{d}_f|A_f^\mathrm{best},\theta_f^\mathrm{best})}{P(\tilde{d}_f|0,0)} \right) \nonumber \\
    &= \sum_{m,n\in\set{\mathrm{real},\mathrm{imag}}}\tilde{d}_f^m\tilde{N}_{mn}^{-1}\tilde{d}_f^n
\end{align}
is the test statistic for the local significance at each frequency. $A_f^\mathrm{best}$ and $\theta^\mathrm{best}$ are the parameters that maximize Eq.~\eqref{eq:Pdf_given_parameters}.

\subsection{Estimation of ALP-photon coupling} \label{sec:estimate_g}
We assume the stochastic ALP field (Eq.~\eqref{eq:ALP_field}) to estimate the ALP-photon coupling, $g_{a\gamma\gamma}$.
We consider the estimator for the square of the ALP photon coupling $g_{a\gamma\gamma}^2$ as
\begin{align}
    \hat{g}_{a\gamma\gamma}^2 
    = \frac{|\tilde{d}_f|^2 - \braket{|\tilde{n}_f|^2}}{F_f^2\phi_\mathrm{DM}^2/4}.
    \label{eq:g2estimator}
\end{align}
and translate it into $g_{a\gamma\gamma}$.
We treat $\hat{g}_{a\gamma\gamma}^2$ as a single parameter and allow it to be negative.
The probability density for $\hat{g}_{a\gamma\gamma}^2$ is constructed by Monte Carlo simulations, using the probability density for obtaining $\tilde{d}_f$
\begin{align}
    P(\tilde{d}_f|g_{a\gamma\gamma}, \phi_\mathrm{DM}) 
    = \frac{\exp\left(-\frac{1}{2}\underset{m,n\in\set{\mathrm{real},\mathrm{imag}}}{\sum}\tilde{d}_f^m\tilde{C}_{f,mn}^{-1}\tilde{d}_f^n \right)}{\sqrt{(2\pi)^2\,\mathrm{det}(\tilde{C}_f)}}
    \label{eq:P_df_g},
\end{align}
where $\tilde{C}_f$ is a covariance matrix of the noise and signal
\begin{align}
    \tilde{C}_f &\equiv \tilde{N}_f + \frac{g_{a\gamma\gamma}^2}{4}F_f^2\tilde{S}_f.
\end{align}
The signal covariance matrix is
\begin{align}
    \tilde{S}_f &\equiv 
    \begin{pmatrix} \phi_\mathrm{DM}^2/2&0\\0&\phi_\mathrm{DM}^2/2
    \end{pmatrix}.
\end{align}
Note that Eq.~\eqref{eq:P_df_g} is obtained by the multivariate Gaussian convolution of the probability density for the local stochastic ALP amplitude
\begin{align}
    P(\tilde{\phi}_f|\phi_\mathrm{DM}) &= \frac{\exp\left(-\frac{1}{2}\underset{m,n\in\set{\mathrm{real},\mathrm{imag}}}{\sum}\tilde{\phi}_f^m\tilde{S}_{f,mn}^{-1}\tilde{\phi}_f^n \right)}{\sqrt{(2\pi)^2\,\mathrm{det}(\tilde{S}_f)}} \\
    &= \frac{1}{\pi\phi_\mathrm{DM}^2}\exp\left(-\frac{\phi_c^2+\phi_s^2}{\phi_\mathrm{DM}^2}\right),
\end{align}
and the probability density for obtaining $\tilde{d}_f$ for given signal parameters (Eq.~\eqref{eq:Pdf_given_parameters}), where we relate the stochastic ALP amplitude $\tilde{\phi}_f$ with $A_f\exp(i\theta_f)$ of Eq.~\eqref{eq:data-signal}.

It is immediately shown that $\hat{g}_{a\gamma\gamma}^2$ is unbiased because the variance of $\tilde{d}_f$ is 
\begin{align}
    \braket{|\tilde{d}_f|^2} &= \braket{|\frac{g_{a\gamma\gamma}}{2}\tilde{\phi}_f + \tilde{n}_f|^2} \nonumber \\
    &= \frac{g_{a\gamma\gamma}^2}{4} \braket{|\tilde{\phi}_f|^2} + \braket{|\tilde{n}_f|^2} \nonumber \\
    &= \frac{g_{a\gamma\gamma}^2}{4} F_f^2\phi_\mathrm{DM}^2 + \braket{|\tilde{n}_f|^2}.
    \label{eq:LSSA_varianve}
\end{align}
As discussed in Appendix~\ref{apx:estimator}, the efficiency (variance) of this estimator is comparable to the estimator used in \citetalias{ALP_PB}, while this estimator is computationally faster.

\section{Data validation}
\label{sec:validation}
To validate the data analysis, we test the internal consistency of the data using a blind analysis framework. Hereafter we call this test as a null test. We follow the similar formalism used in \citetalias{ALP_PB}.
We split the data in half and form null statistics by taking a difference to cancel the oscillation signal but amplify the systematic errors, and analyze the consistency of the null statistics with the null hypothesis.
The data-splits are largely categorized into two types of splits to probe for different type of systematics.

\textbf{Bolometer-splits}: These splits probe the mis-calibration between detectors or the systematic error in each observation such as the time drift of the detector characteristics over one hour.

\textbf{Observation-splits}: These splits probe the systematic variations between observations.

In addition to the data-splits considered in \citetalias{PB20}, we introduce the following data splits dedicated to probe the systematics of the polarization angle.
\begin{enumerate}
    \setlength{\itemsep}{0pt}
    \setlength{\parskip}{0pt}
    \setlength{\parsep}{0pt}
    \item 2f/4f angle by obs/bolo: These split the data into observations or detectors by angle of signal synchronous with the rotation of the HWP, to probe the detector-by-detector or observation-by-observation mis-calibration of the HWP angle. 
    \item Time constant by bolo: This splits the data into detectors with larger time constants and those with smaller time constants, to probe the mis-calibration of detector's time constants.
\end{enumerate}

\subsection{Null statistics}
We construct three different types of null statistics.
First, we take noise-weighted differences as 
\begin{align}
    \chi^\mathrm{DC}_\mathrm{null} = \frac{\sum\limits_{t\ \in\ \mathrm{split0}}{d_{t,0}\sigma_{t,0}^{-2}}}{\sum\limits_{t\ \in\ \mathrm{split0}}\sigma_{t,0}^{-2}} 
    - \frac{\sum\limits_{t\ \in\ \mathrm{split1}}{d_{t,1}\sigma_{t,1}^{-2}}}{\sum\limits_{t\ \in\ \mathrm{split1}}\sigma_{t,1}^{-2}},
    \label{eq:chi_DC}
\end{align}
where $d_{t,k}$ and $\sigma_{t,k}$ are the $k$-th split data and its error bar.
For the data-splits which split the data periodically, we did not use $\chi^\mathrm{DC}_\mathrm{null}$ for null test because of the risk of unblinding the signal (Table. \ref{table:nulltest_individual}).
The other null statistics are constructed in the frequency domain, by the difference of LSSA for each data-split as
\begin{align}
    \chi^\mathrm{AC}_\mathrm{null} &= (\tilde{d}^m_{f,0}-\tilde{d}^m_{f,1})/\tilde{D}_{mm}, \\
    \chi^2_\mathrm{null} &= \sum_{m,n\in\set{\mathrm{real, imag}}} (\tilde{d}^m_{f,0}-\tilde{d}^m_{f,1})\tilde{D}^{-1}_{mn}(\tilde{d}^n_{f,0}-\tilde{d}^n_{f,1}),
    \label{eq:chi_AC}
\end{align}
where $\tilde{d}_{f,k} = \tilde{d}_{f,k}^\mathrm{real} + i\tilde{d}_{f,k}^\mathrm{imag}$ is the LSSA of $k$-th split data (Eq.~\eqref{eq:LSSA_complex_matrix}) and $\tilde{D}_{mn}$ is the covariance matrix of the difference of LSSAs.
$\chi^\mathrm{DC}_\mathrm{null}$ and $\chi^\mathrm{AC}_\mathrm{null}$ follows the chi-square distribution with one degree of freedom, $\chi^2_\mathrm{null}$ approximately follows the chi-square distribution with two degrees of freedom.

As discussed in Sec.~\ref{sec:spectrum}, the null statistics for different frequency bins are correlated. 
The pattern of correlation differs for the observation-splits, which introduces imperfect signal cancellation in the null statistics for the observation-splits. 
This is especially problematic when large sinusoidal signals are present, but is sufficiently small in this study since we are verifying the presence of a sinusoidal signal, which is known to be small. For this reason, we do not apply the transfer function (Sec.~\ref{sec:spectrum}) for the construction of null statistics (Eq.~\eqref{eq:chi_AC}).

\subsection{Null test results} \label{sec:nulltest_result}
\begin{figure}
    \centering
    \includegraphics[width = 0.49\textwidth]{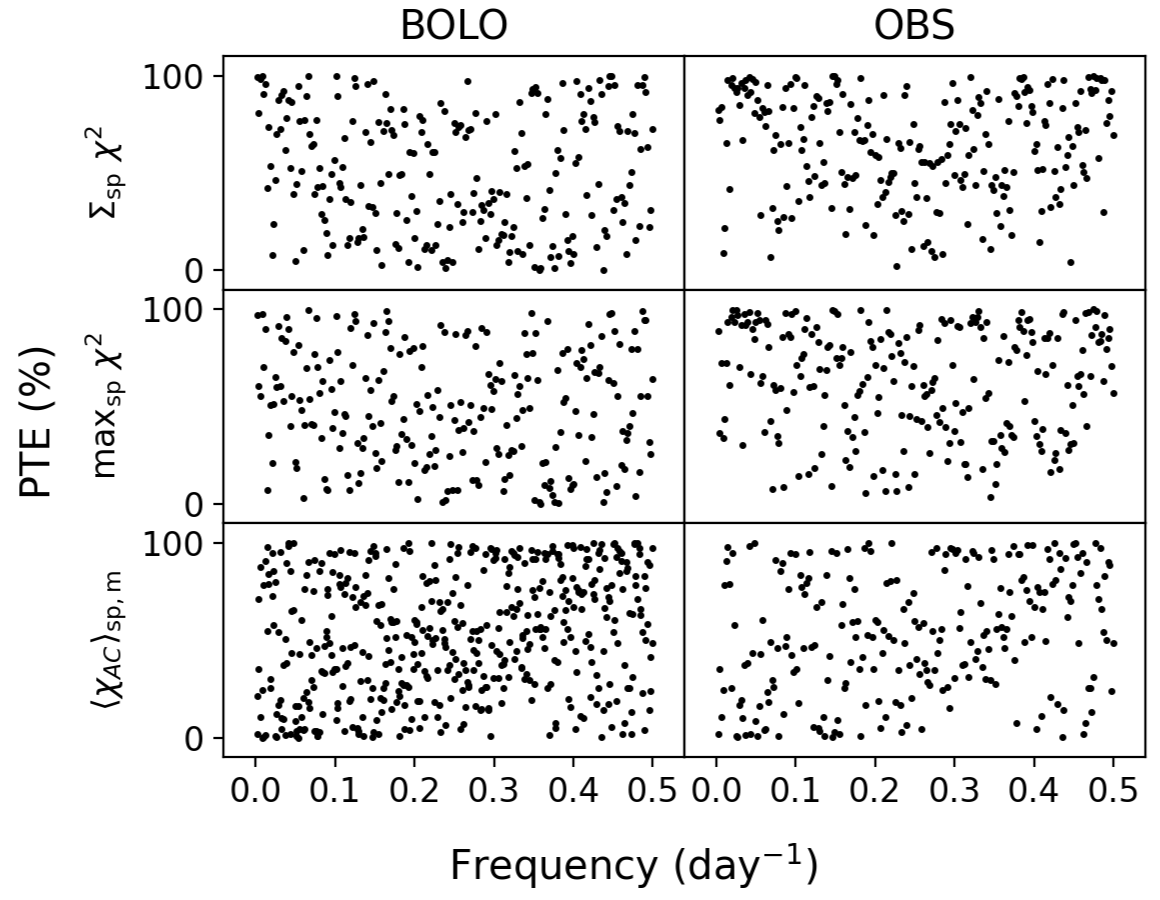}
    \caption{PTEs of null statistics for individual frequency bins. The subscript ``sp" denotes ``data-splits." Although there are correlations between frequency bins, there is no significant deviation from a uniform distribution.}
    \label{fig:null_spectrum}
\end{figure}

Table.~\ref{table:nulltest_individual} shows the null statistics for each data-splits.
The null statistics are compared with the 1440 signflip noise only simulations\footnote{48 samples from each of the 30 signflip maps.} (Sec.~\ref{sec:map2angle}). Then we evaluate the probability to exceed value (PTE) by counting the number of simulation whose null statistics exceeds the real data. 
Table~\ref{table:nulltest_individual} shows the distribution of PTE for each data split, and Fig.~\ref{fig:null_spectrum} shows the distribution of PTE for each frequency bin. 
\begin{table}
    \caption{PTEs of null statistics for individual data-splits}
    \label{table:nulltest_individual}
    \centering
    \begin{tabular*}{.49\textwidth}{@{\extracolsep{\fill}}lcccc}
    \hline \hline
    Data split & \multicolumn{4}{c}{PTE (\%)} \\
    & $\underset{f}{\operatorname{\sum}}\chi^2_\mathrm{null}$ & $\underset{f}{\operatorname{\mathrm{max}}}\chi^2_\mathrm{null}$ & $\braket{\chi^\mathrm{AC}_\mathrm{null}}_{f,m}$&$\chi^\mathrm{DC}_\mathrm{null}$\\
    \hline
    \textbf{Observing conditions} \\
    4th vs. 5th season$^*$ & 14.2 & 7.2 & 31.5 & 95.3$^{\dag}$ \\
    High vs. low PWV$^*$ & 12.6 & 53.2 & 64.4 & 15.8 \\
    Sun distance$^*$ & 11.4 & 4.2 & 20.8 & 15.8$^{\dag}$ \\
    Moon distance$^*$ & 13.0 & 48.0 & 79.5 & 15.8$^{\dag}$ \\
    Left vs. right subscans & 50.1 & 59.4 & 59.7 & 67.1 \\
    \hline
    \textbf{Instrument} \\
    Top vs. bottom bolo & 12.0 & 18.8 & 76.6 & 86.7\\
    Q vs. U pixels & 61.5 & 26.0 & 30.4 & 56.8 \\
    Top vs. bottom half & 11.3 & 11.8 & 49.5 & 34.3 \\
    Left vs. right half & 52.6 & 71.4 & 16.9 & 67.1\\
    \hline
    \textbf{Data quality} \\
    I2P leakage by obs$^*$ & 12.2 & 13.5 & 5.1 & 64.8 \\
    I2P leakage by bolo & 24.0 & 33.3 & 65.4 & 85.5 \\
    Common mode Q knee$^*$ & 11.9 & 56.5 & 43.0 & 48.5 \\
    Common mode U knee$^*$ & 14.4 & 22.6 & 18.5 & 16.5 \\
    Common noise by obs$^*$ & 11.1 & 8.8 & 77.1 & 59.1 \\
    Common noise by bolo & 98.5 & 84.7 & 15.3 & 46.5 \\
    2f amplitude by obs$^*$ & 11.2 & 11.7 & 13.5 & 10.0 \\
    2f amplitude by bolo & 30.1 & 41.2 & 68.4 & 39.2 \\
    4f amplitude by obs$^*$ & 11.2 & 11.7 & 13.5 & 37.6 \\
    4f amplitude by bolo & 30.1 & 41.2 & 68.4 & 39.2 \\
    2f angle by obs$^*$ & 13.3 & 50.8 & 1.9 & 0.6 \\
    2f angle by bolo & 78.4 & 57.7 & 15.2 & 33.5 \\
    4f angle by obs$^*$ & 12.6 & 4.8 & 14.7 & 27.8 \\
    4f angle by bolo & 57.3 & 30.3 & 14.5 & 94.7 \\
    Gain by obs$^*$ & 11.2 & 11.7 & 13.5 & 38.1 \\
    Tau by bolo & 16.5 & 28.3 & 57.1 & 65.8 \\
    \hline \hline
    \end{tabular*}
    \vspace{0ex}
     {\raggedright \textbf{NOTE}: $^*$ indicates observation-split type of the data split. $\chi^\mathrm{DC}_\mathrm{null}$ with $^\dag$ were blinded during the data validation process, because these partially unblind oscillation signal at certain frequencies. \par}
\end{table}

Next, we compute six representative test statistics for bolometer-splits and observation-splits separately:
(1) the average of $\chi^\mathrm{DC}_\mathrm{null}$ among all data-splits, 
(2) the average of $\chi^\mathrm{AC}_\mathrm{null}$ among all data-splits, 
(3) the most extreme total $\chi^2_\mathrm{null}$ by data-splits summed over all frequencies, 
(4) the most extreme total $\chi^2_\mathrm{null}$ by all frequencies summed over all data-splits, 
(5) the most extreme total $\chi^2_\mathrm{null}$ among all frequencies and all data-splits,
(6) the total $\chi^2_\mathrm{null}$ summed over all frequencies and all data-splits.
(1) and (2) probe for the biases, (3)--(5) probe for the outliers specific to the data-splits or frequencies, and (6) probes for the mis-estimation of our uncertainties.
The PTEs for the representative test statistics are also computed by comparison with noise only simulations.

As a pass criteria of the null test, we require the PTE of the lowest PTE value to be larger than 5\% to probe the biases and outliers. 
We also require the PTE of the highest PTE value\footnote{The PTE of the lowest (highest) PTE value is obtained by counting the number of simulations whose lowest (highest) PTE value is smaller (larger) than the observed lowest (highest) PTE value.} to be larger than 5\% to check the mismatch between the real and estimated uncertainties.\footnote{The Kolmogorov-Smirnov (KS) test \citep{KStest1951} is often used to check the mismatch between the true and estimated uncertainties. However, the KS test is not a strong statistical test for this study, because the null statistics of this study are correlated, while the KS test assumes independent and identically distributed samples of test statistics.}
The numerical values for these statistics are given in Table~\ref{table:nulltest_summary}. 
Both bolometer-split and observation-split pass the criteria.

\begin{table}
    \caption{PTEs of representative test statistics used for null test pass criteria.}
    \label{table:nulltest_summary}
    \centering
    \begin{tabular*}{.45\textwidth}{@{\extracolsep{\fill}}lcc}
    \hline \hline
    Type of test &\multicolumn{2}{c}{PTE (\%)} \\
    & BOLO & OBS \\
    \hline
    Average $\chi^\mathrm{DC}_\mathrm{null}$ overall & 91.46 & 11.04 \\
    Average $\chi^\mathrm{AC}_\mathrm{null}$ overall & 52.08 & 12.78 \\
    Extreme $\chi^2_\mathrm{null}$ overall & 65.49 & 12.43 \\
    Extreme $\chi^2_\mathrm{null}$ by split & 11.74 & 16.67 \\
    Extreme $\chi^2_\mathrm{null}$ by frequency & 54.37 & 16.18 \\
    Total $\chi^2_\mathrm{null}$ & 36.60 & 14.37 \\
    \hline
    Lowest p-value & 40.90 & 35.42 \\
    Highest p-value & 34.47 & 99.68 \\
    \hline \hline
    \end{tabular*}
\end{table}

\subsection{Correlation with systematic template} \label{sec:syst_correlation}
In order to detect possible systematic errors,  
we additionally test if any significant correlation coefficient is found between the measured Tau~A polarization angle and the variation of instrumental conditions or environmental conditions, which could potentially produce systematic time-dependent polarization angle fluctuation.
We call the representative fluctuation of instrumental or environmental conditions used in this systematic test as systematic templates. 
The systematic templates are constructed using the \textsc{Polarbear}'s instrumental or environmental monitors. We perform this systematic test using the following systematic templates: the average of PWV, the ambient temperature, the focal plane stage temperature, and the polarization angle of the mirror polarization.\footnote{We consider that revealing the correlation coefficient does not violate the blind policy. This is because the systematic template contains no signal, and the correlation coefficient is only sensitive to the possible systematic contamination, and is insensitive to the variability of Tau~A or axion-like oscillations.}

\begin{figure}
    \centering
    \includegraphics[width = .47\textwidth]{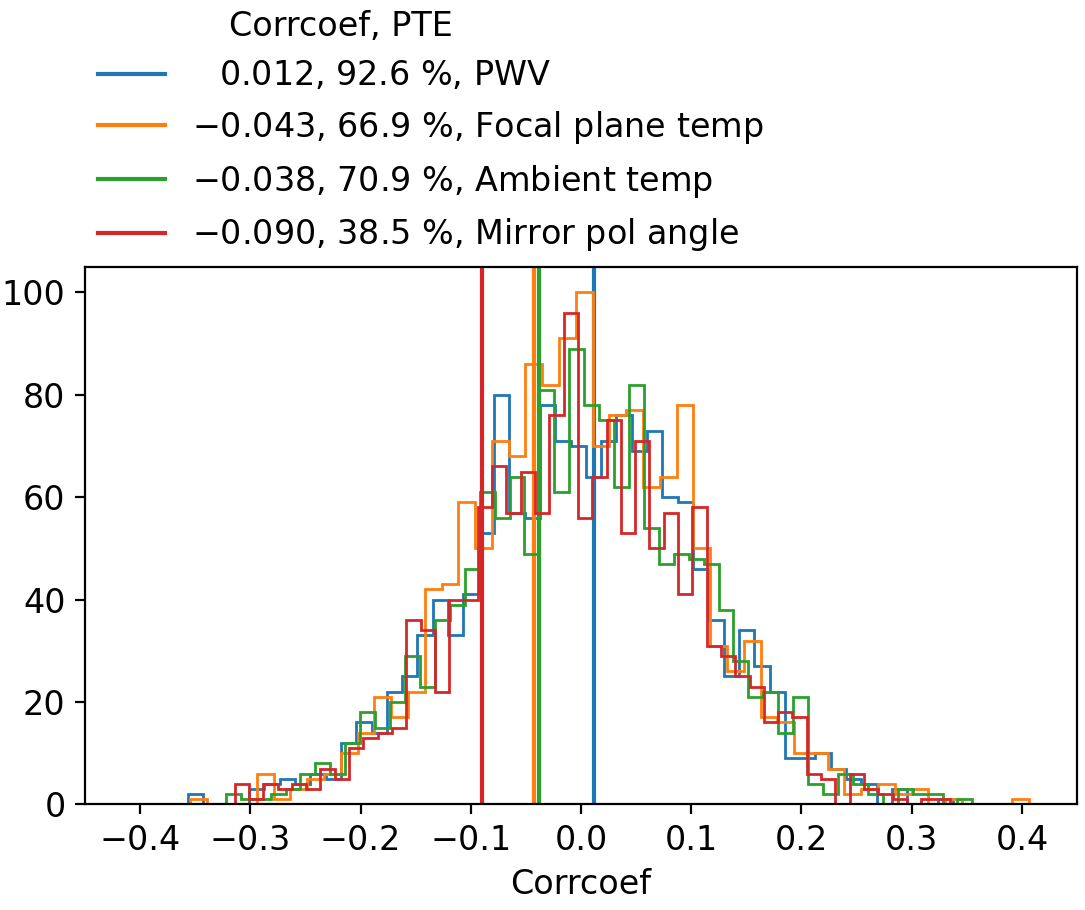}
    \caption{Correlation coefficient between Tau~A angle and systematic templates. The vertical lines represent correlation coefficient between measured Tau~A angle and systematic templates. The histograms represent correlation coefficient between the noise simulations and systematic templates.}
    \label{fig:corrcoef_test}
\end{figure}

We define the weighted correlation coefficient between data $d_t$ with error bar $\sigma_t$ and the systematic template $T_t$ as
\begin{align}
    \mathrm{Corrcoef}(d, T) = 
    \frac{\mathrm{Cov}(d, T)}{\sqrt{\mathrm{Cov}(d, d)~\mathrm{Cov}(T, T)}},
    \label{eq:corrcoeff}
\end{align}
where  
\begin{align}
    \mathrm{Cov}(d, T) = 
    \frac{\sum_t (d_t - \braket{d})(T_t - \braket{T}) \sigma_t^{-2}}{\sum_t \sigma_t^{-2}}
\end{align}
is the weighted covariance, and
\begin{align}
    \braket{d} = \frac{\sum_t d_t \sigma_t^{-2}}{\sum_t \sigma_t^{-2}}, \braket{T} = \frac{\sum_t T_t \sigma_t^{-2}}{\sum_t \sigma_t^{-2}}
\end{align}
are the weighted average.
Figure~\ref{fig:corrcoef_test} shows the significance of the correlation coefficients obtained by comparing the correlation coefficients for 1440 noise realizations, and no significant correlation coefficient is found.

Earlier in the data analysis, this method was beneficial to identify systematics and determine the optimum calibration and analysis strategies.
This method led to the detection of the PWV-dependent mis-estimation of the I2P leakage and the mis-calibration of the polarization angle in the telescope coordinate with correlation coefficients 0.51 and 0.55, respectively, resulting in the improvement of the angle calibration discussed in Sec~\ref{sec:calibration}.

\section{Results} \label{sec:results}
We report the results of Tau~A polarization angle, its frequency spectrum, and upper bounds of its oscillation amplitude. We also report the upper bounds of ALP-photon coupling, assuming that no sources other than ALP are causing Tau~A's polarization angle variation.

\begin{figure*}
    \centering
    \includegraphics[width = 0.98\textwidth]{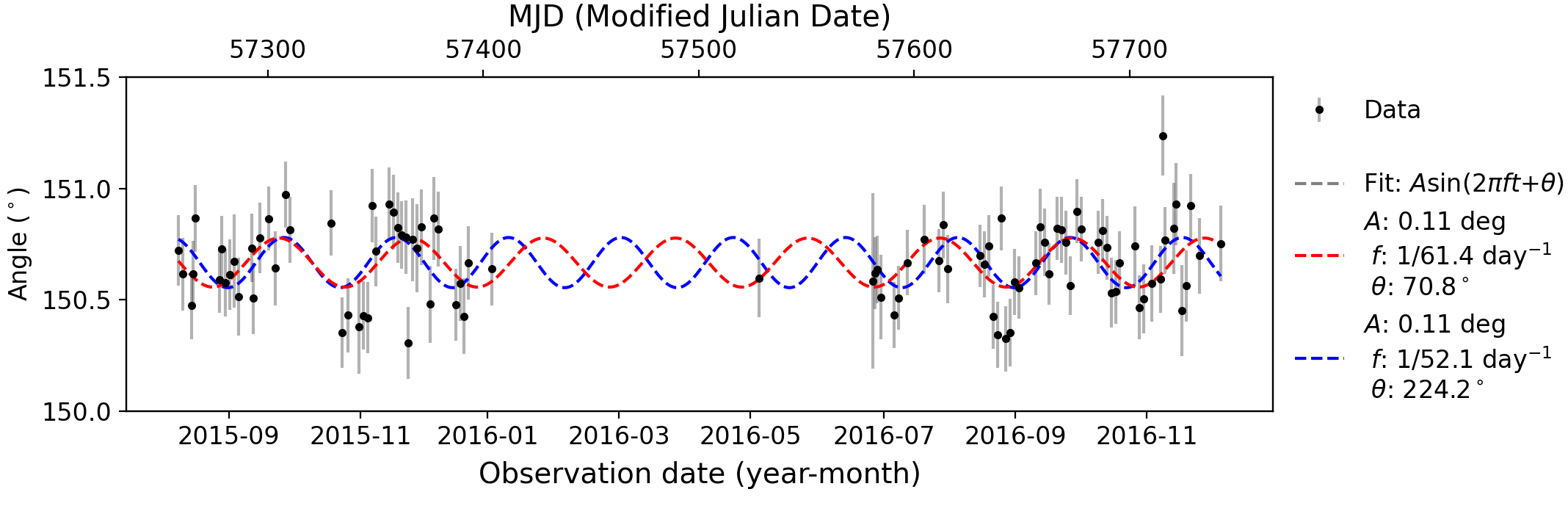}\par
    \includegraphics[width = 0.98\textwidth]{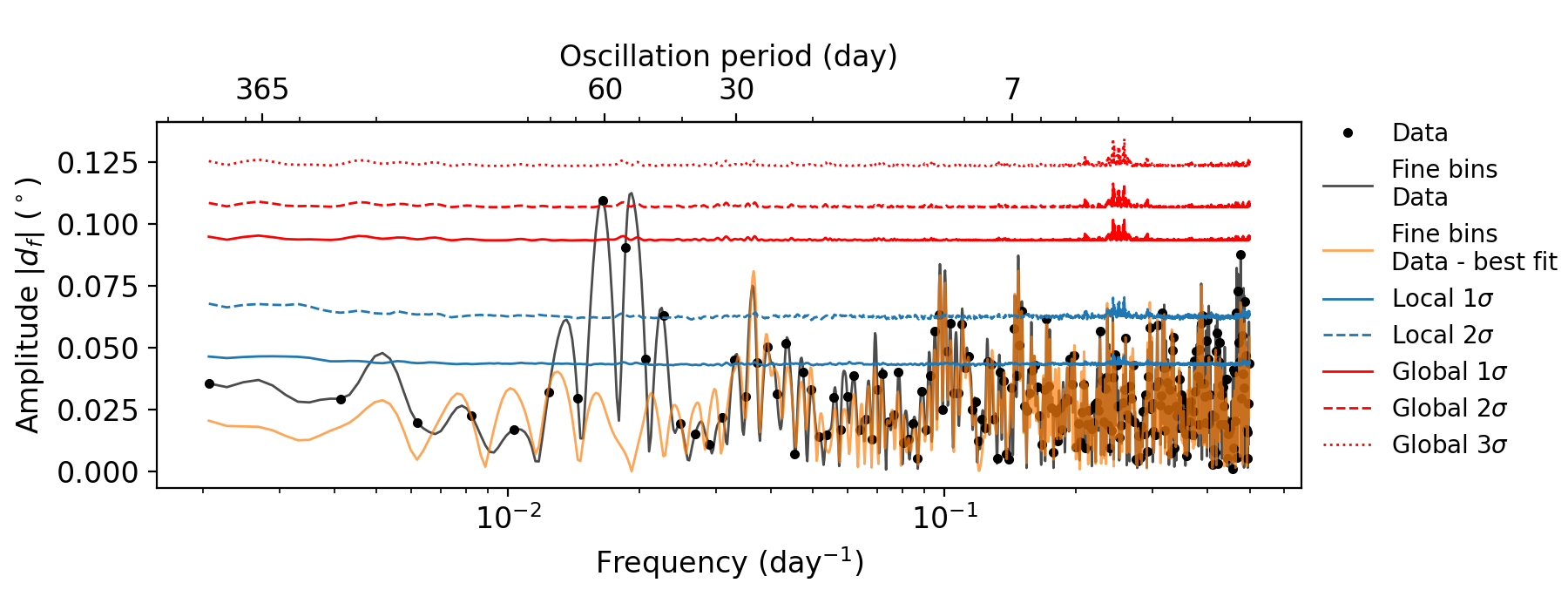}
    \caption{Top: The measured polarization angle of Tau~A superimposed on the best-fit sine curves. 
    The phase of the oscillation ($\theta$) is defined in MJD. 
    Number of observations is small from January 2016 to July 2016, because angular distance between Sun and Tau~A is small in this period, also because of the maintenance of the instruments. 
    There are two equally significant correlated oscillation modes at 1/61\,day$^{-1}$ and 1/52\,day$^{-1}$. 
    Bottom: Amplitudes of the LSSA of data compared with the approximated amplitudes with certain local and global significance levels. 
    The black dots shows the LSSA of data at the frequency bins of Eq.~\eqref{eq:fbins}, the black line shows the LSSA of data at 10 times more finely sampled frequency bins. 
    The orange solid line shows the LSSA of data minus the best fit sine curve at 1/52\,day$^{-1}$.
    }
    \label{fig:data}
\end{figure*}
\begin{figure}
    \centering
    \includegraphics[width = 0.47\textwidth]{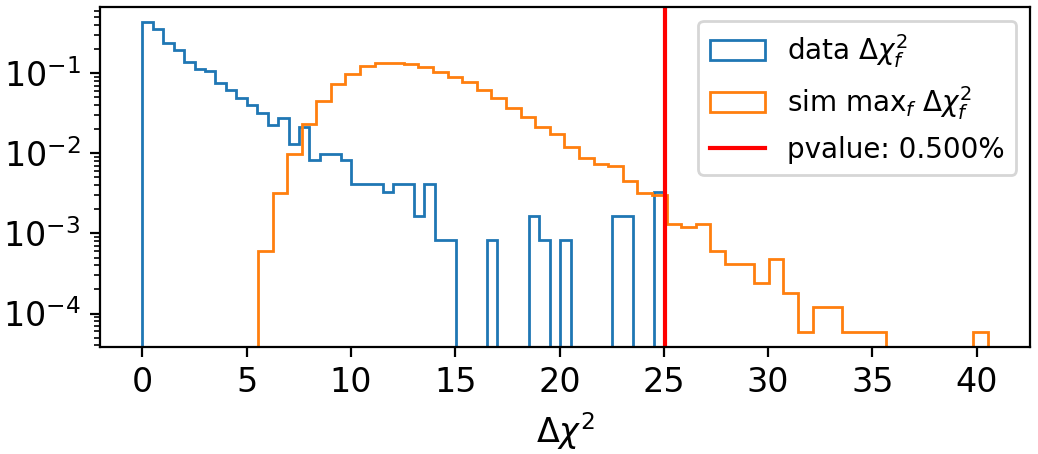}
    \caption{The global significance of oscillation mode at 61\,day$^{-1}$ is 2.5$\sigma$, corresponding to a p-value of 0.5\%. The red lines shows the global significance.}
    \label{fig:significance}  
\end{figure}

Our estimated timestream of Tau~A polarization angle is shown in the top panel of Fig.~\ref{fig:data}. 
The black dots in the bottom panel of Fig.~\ref{fig:data} shows its LSSA at the frequency bins below $f_\mathrm{NYQ}$ (Eq.~\eqref{eq:fbins}).
To evaluate the detection significance, test statistics were computed using a total of 24,000 noise realizations generated based on the bootstrap method (Sec.~\ref{sec:map2angle}) to have sufficient accuracy of the significance level up to 3$\sigma$.
The blue lines and red lines in the bottom panel of Fig.~\ref{fig:data} show the approximated amplitudes with certain local and global significance levels.\footnote{For example, the amplitude of 1$\sigma$ local significance level is $\sqrt{\mathrm{percentile}(\Delta \chi^2_f, 84.1)\braket{|n_f|^2}}$ and the amplitude of 1$\sigma$ global significance level is $\sqrt{\mathrm{percentile}(\Delta \chi^2, 84.1)\braket{|n_f|^2}}$.}
Figure \ref{fig:significance} shows the distribution of test statistics.
The largest significance level we found is p-value of 0.50\%, corresponding to 2.5$\sigma$, at 1/61\,day$^{-1}$. 

The black solid line in Fig.~\ref{fig:data} shows the LSSA with 10 times finer frequency bins than Eq~\eqref{eq:fbins}.
Another peak at 1/52\,day$^{-1}$, which is not captured by the sparse frequency bins, is found to have a global significance level of 2.5$\sigma$ by calculating the test statistics over the 10 times finer frequency bins. We confirmed that the null-test pass criteria were still met with 10 times finer frequency bins. The local significance levels of these two most significant oscillation modes at 1/61\,day$^{-1}$ and 1/52\,day$^{-1}$ are found to be 4.4$\sigma$ and 4.5$\sigma$ analytically. We note that the local significance is not relevant as a detection significance because we are examining many frequency bins.

The orange solid line in Fig.~\ref{fig:data} shows the LSSA of data minus the best fit sine curve at 1/52\,day$^{-1}$.
The data minus best fit is consistent with the null hypothesis, with the global significance level smaller than 1$\sigma$. 
This means that the two most significant oscillation modes are correlated, and the result can be interpreted as a hint of a single-mode oscillation signal. 

If we interpret these two most significant oscillations as axion-like polarization signals, then the corresponding ALP masses are \SI{7.8e-22}{eV} and \SI{9.2e-22}{eV}, respectively. The possible interpretations of the hint of a signal are discussed in Sec.~\ref{sec:discussion}.

\begin{figure}
    \centering
    \includegraphics[width = 0.49\textwidth]{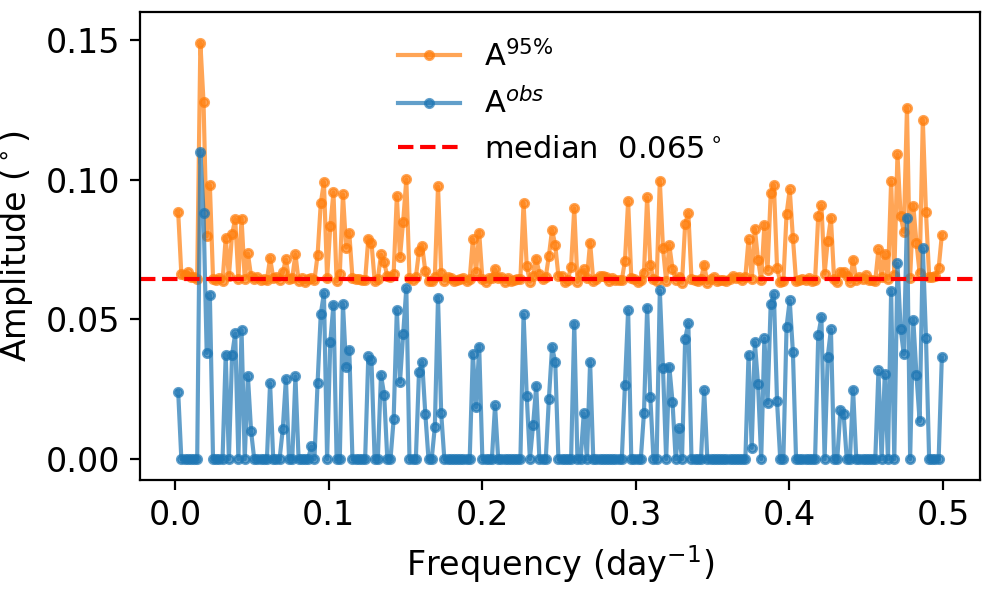}\par
    \includegraphics[width = 0.49\textwidth]{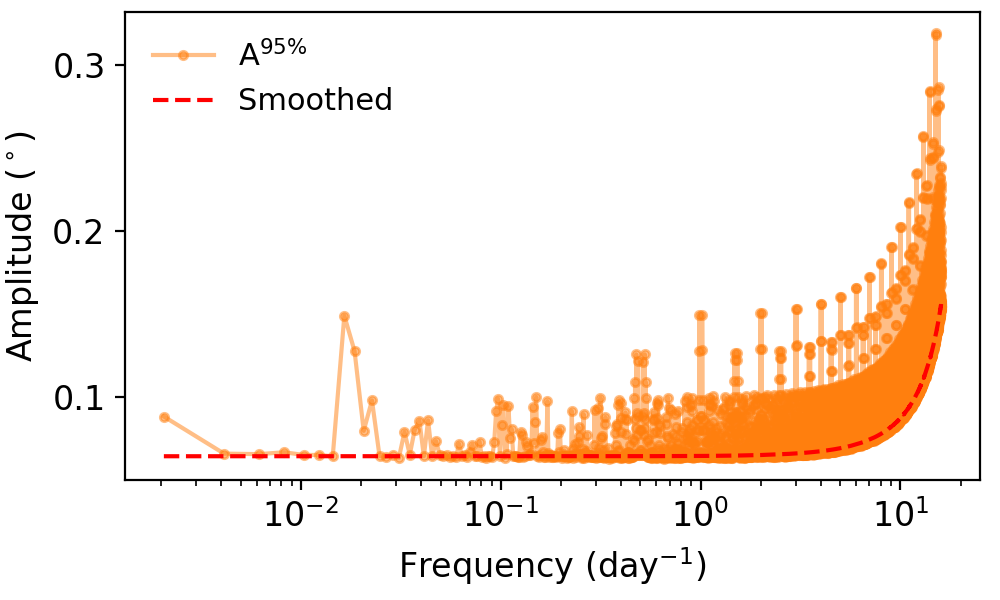}
    \caption{Top: 95\% upper bound and measured polarization oscillation amplitudes below $f_\mathrm{NYQ}$. Bottom: Extended 95\% upper bound of polarization oscillation amplitudes.}
    \label{fig:A_upperlimit}
\end{figure}

We report 95\% upper bounds on the oscillation amplitudes and $g_{a\gamma\gamma}$, because no point exceeds the significance level of 3$\sigma$. 
The 95\% upper bounds on the oscillation amplitudes at the frequency bins below $f_\mathrm{NYQ}$ are shown in the top panel of Fig.~\ref{fig:A_upperlimit}. 
As a measure of our experimental sensitivity, we report the median upper bound below $f_\mathrm{NYQ}$ as 0.065$^\circ$.
The upper bound $A^{95\%}$ is obtained with the frequentist approach by the classic Neyman construction \citep{Neyman:1937uhy} as 
\begin{align}
    \int_0^{\hat{A}_\mathrm{obs}}d\hat{A}~P(\hat{A} | A^{95\%}) = 0.05,
\end{align}
where $\hat{A}$ is the maximum likelihood estimator of $A_f$ using Eq.~\eqref{eq:A_likelihood}. $P(\hat{A}|A)$ is the probability density for $\hat{A}$, which is obtained by Monte Carlo simulations of $\hat{A}$ using Eqs.~\eqref{eq:A_likelihood} and \eqref{eq:Pdf_given_parameters}. 
The bottom panel of Fig.~\ref{fig:A_upperlimit} shows the 95\% upper bounds extended to frequency bins larger than $f_\mathrm{NYQ}$ (Eq.~\eqref{eq:fbins_science}).
Due to aliasing, the signal seen at $f$ ($<f_\mathrm{NYQ}$) may be aliased from higher frequencies.
The red dashed curve represents smoothed approximation of our upper bounds
\begin{align}
    A < (0.065^\circ) \times \left[\mathrm{sinc}\left(\frac{f}{\SI{7.6}{day^{-1}}}\right)\right]^{-1}, \label{eq:median_A}
\end{align}
where sinc function approximates the transfer function $F_f$ (Eq.~\eqref{eq:smoothed_transfer}).

\begin{figure*}
    \centering
    \includegraphics[width = 0.98\textwidth]{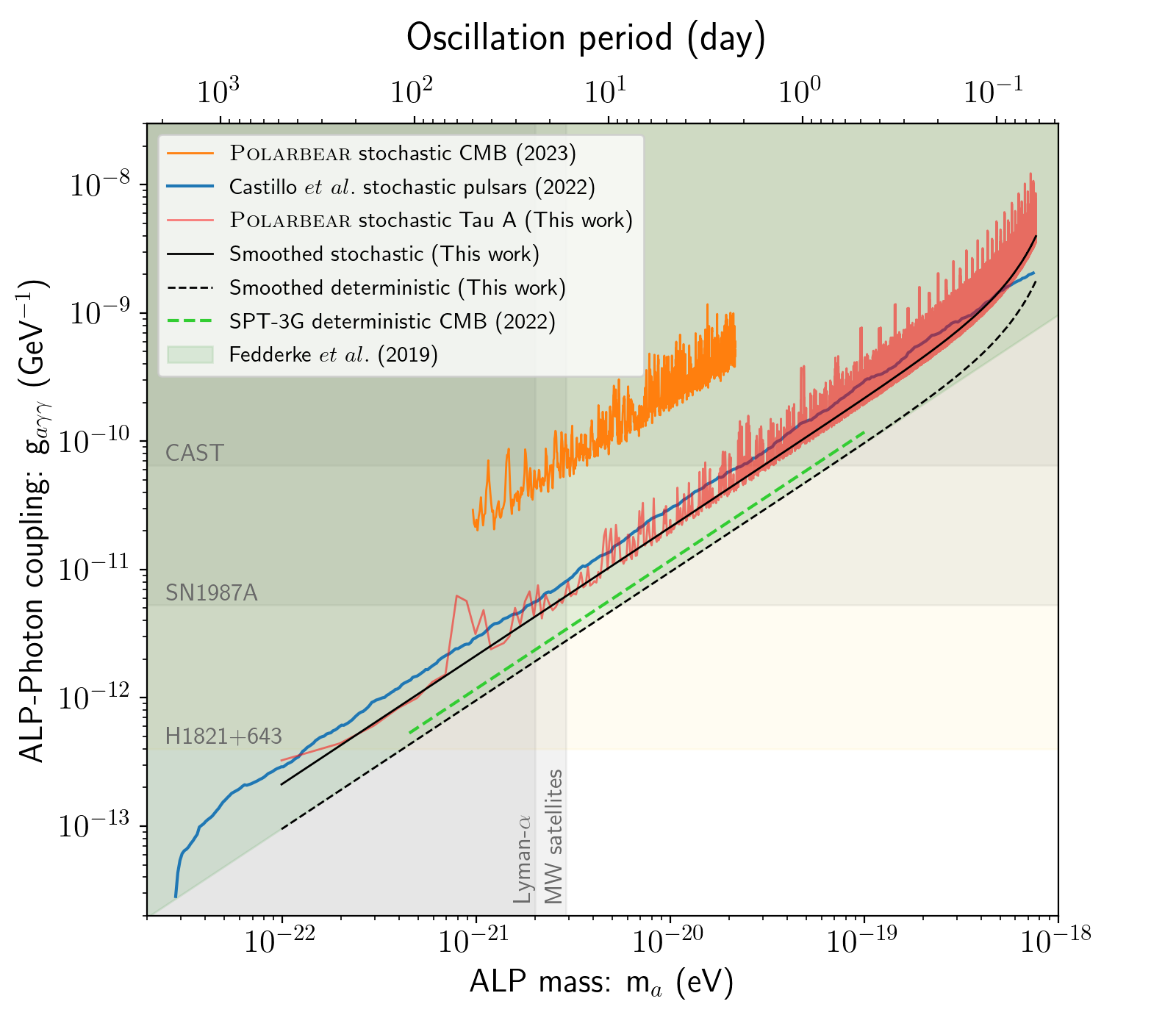}
    \caption{95\% Upper bound of $g_{a\gamma\gamma}$. The upper bounds assuming a stochastic ALP field is shown using solid lines. 
    The orange solid curve represents a constraint by searching for an axion-like oscillation from CMB \citepalias{ALP_PB}. 
    The blue solid curve represents a constraint by searching for an axion-like oscillation from various pulsars and Tau~A \citep{Castillo_2022}. 
    The red solid curve represents our stochastic bounds, the primary results of this study. The black curve represents the smoothed approximation of our stochastic bounds (Eq.~\eqref{eq:median_stoch}).
    The upper bounds assuming deterministic ALP field are shown using dashed curves. The green dashed line represents a constraint by searching for an axion-like oscillation of CMB \citep{ALP_SPT3g}.
    The black dashed curve represents the smoothed approximation of our deterministic bounds (Eq.~\eqref{eq:median_det}).
    The green shaded bound represents a constraint from the absence of the suppression of CMB polarization due to an axion-like oscillation in the recombination era \citep{Fedderke2019}. The upper bound from the CAST experiment \citep{cast_2017}, an absence of the gamma-ray excess of SN1987A \citep{Payez:2014xsa}, and an absence of the X-ray spectral distortions of the quasar H1821+643 \citep{Chandra2022} are shown. 
    The lower bound on the ALP mass by the Lyman-$\alpha$ forest \citep{lyman_alpha_fuzzy} and the Milky Way satellite galaxies \citep{Nadler2021} are also shown.}
    \label{fig:constraint_on_g}
\end{figure*}

The 95\% upper bounds of $g_{a\gamma\gamma}$ assuming the stochastic ALP field are also constructed using the Neyman construction. 
We consider the constraint on the square of the effective amplitude of polarization oscillation, $g^2\equiv g_{a\gamma\gamma}^2 F_f^2 \phi^2_\mathrm{DM}/4$, and convert it to a constraint on $g_{a\gamma\gamma}$. 
The detailed derivation of it is described in Appendix~\ref{apx:bound}.\footnote{The stochastic bounds obtained by this method are 2.2 times conservative than the deterministic bounds using Bayesian inference with a flat prior. The scaling between these two are consistent with our previous result demonstrated in \citetalias{ALP_PB}.}
The median 95\% upper bound on the effective oscillation amplitudes below $f_\mathrm{NYQ}$ is $\sqrt{g^{2,95\%}} = 0.13^\circ$.
Assuming that ALP constitutes all the local dark matter ($\kappa=1$, $\rho_0=\SI{0.3}{GeV/cm^3}$), the smoothed approximation of our stochastic bounds is
\begin{align}
    \begin{split}
    g_{a\gamma\gamma} \leq (\SI{2.16e-12}{GeV^{-1}}) \times\left(\frac{m_a}{10^{-21}\,\mathrm{eV}}\right)\\
    \quad\times\left[\mathrm{sinc}\left(\frac{m_a}{\SI{3.7e-19}{eV}}\right)\right]^{-1}.
    \end{split}
    \label{eq:median_stoch}
\end{align}
Our bounds are shown in Fig.~\ref{fig:constraint_on_g} along with the other selected constraints.
Our bounds for individual frequencies and the smoothed approximation are shown using red and black curves, respectively.
The black dashed line shows the 95\% upper bounds assuming that the amplitude of the ALP field is deterministically given by the density of local dark matter, so called ``deterministic bounds."
Our smoothed approximation of our deterministic bounds is\footnote{The deterministic bounds are obtained by substituting $A^{95\%}$ as the effective amplitude $\sqrt{g^2}$ in Eq.~\eqref{eq:upperbound}.}
\begin{align}
    \begin{split}
    g_{a\gamma\gamma}^\mathrm{deterministic} \leq (\SI{1.0e-12}{GeV^{-1}}) \times\left(\frac{m_a}{10^{-21}\,\mathrm{eV}}\right)\\
    \quad\times\left[\mathrm{sinc}\left(\frac{m_a}{\SI{3.7e-19}{eV}}\right)\right]^{-1}.
    \end{split}
    \label{eq:median_det}
\end{align}
Our primary science result is the stochastic bounds, because the deterministic bounds do not account for the expected variation of the ALP field.

\section{Systematic errors} \label{sec:systematics}
\begin{figure*}
    \centering
    \includegraphics[width = \textwidth]{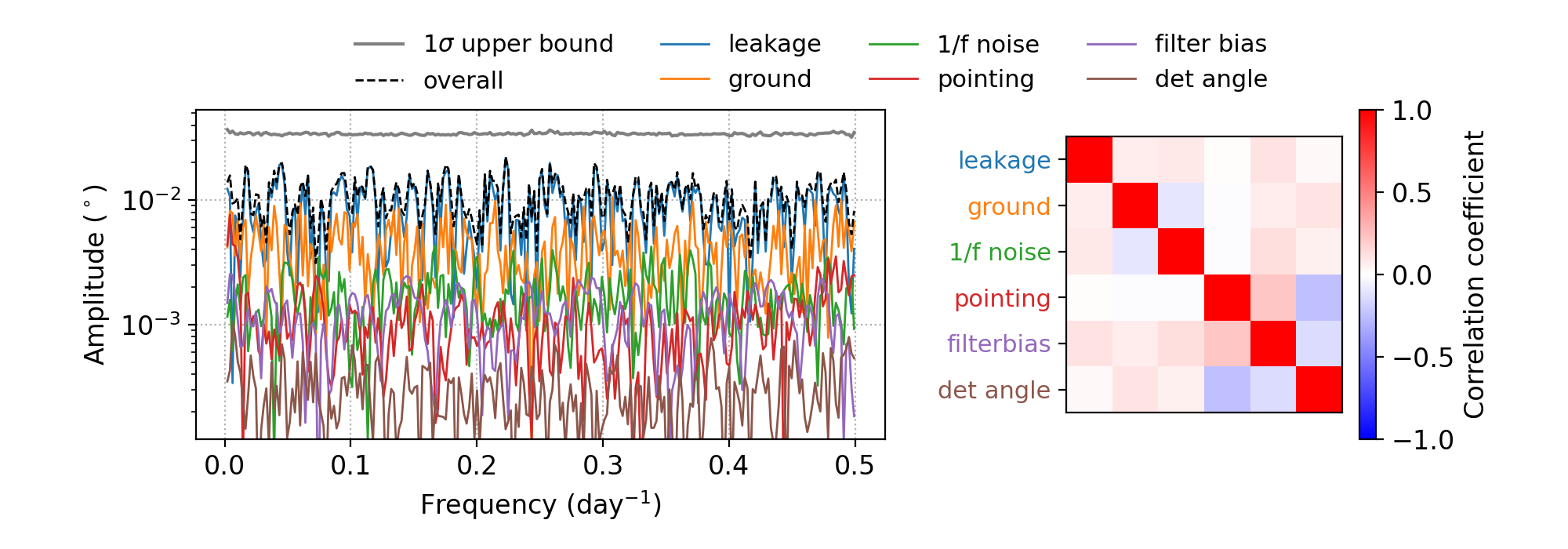}
    \caption{1$\sigma$ upper bound of the statistical uncertainty compared with the systematics. The overall systematics are formed by quadrature sum of individual systematics. The right panel shows the correlation coefficients between different systematics which are smaller than 0.3.}
    \label{fig:systematics_spectrum}
\end{figure*}

This section presents the dedicated studies for the evaluation of possible systematic errors. 
Table~\ref{table:syst} shows the median statistical uncertainty per observation, and the RMS of the systematics, the systematic time-dependent polarization angle fluctuations.
\begin{table}
    \caption{Error budget of the polarization angle of Tau~A. 
    The median statistical uncertainty per observation and the RMS of the systematic polarization angle fluctuations are shown.
    }
    \label{table:syst}
    \centering
    \begin{tabular*}{.45\textwidth}{@{\extracolsep{\fill}}lc}
    \hline \hline
    Type of statistical error & Median $\sigma_t$ (deg)\\
    \hline
    Statistical error of Tau~A & 0.16 \\ 
    Polarization angle calibration by $A_4$ & 0.02 \\ 
    \hline \hline\\
    \hline \hline
    Type of systematic error & RMS (deg) \\
    \hline
    I2P leakage & 0.06 \\ 
    Ground & 0.03 \\ 
    Residual $1/f$ noise & 0.01 \\ 
    MD breaking & $<$ 0.01 \\ 
    Pointing & 0.006  \\ 
    Time domain filter bias & 0.005 \\ 
    Polarization angle of detectors & 0.002 \\ 
    \hline \hline
    \end{tabular*}
\end{table}
The frequency distribution of systematic polarization angle fluctuation is estimated by its LSSA. 
The overall systematics as well as the individual systematics are shown in Fig.~\ref{fig:systematics_spectrum}. 
The systematics are shown up to $f_\mathrm{NYQ}$, because systematics and statistical error scales the same at frequencies higher than $f_\mathrm{NYQ}$. 
The correlation coefficient between different systematics are found to be smaller than 0.3, as shown in the right panel of Fig.~\ref{fig:systematics_spectrum}. 
The overall systematics are formed by quadrature sum of individual systematics.
We find the total systematics to be subdominant compared to the statistical error at all frequencies.
Possible residual systematic errors are discussed in Sec~\ref{sec:discussion}.

\subsection{Polarization angle calibration} \label{sec:syst_HWPangle}
The uncertainty of polarization angle calibration in the telescope coordinate produces additional statistical uncertainty in the measured Tau~A polarization angle. 
The overall uncertainty of the polarization angle calibration in telescope coordinate is given by the statistical error of the instrumental polarization angle per observation, 0.02$^\circ$ STD (Sec.~\ref{sec:polcalobs}).
This uncertainty degenerately includes time constant, timing offset, and HWP angle uncertainties. 
The upper bound of the uncertainty due to the time constant and timing offset is estimated independently for the cross check from the chopped thermal source calibration. These are found to be $< 0.04^\circ$ and $< 0.03^\circ$, respectively. 

\subsection{Intensity to polarization leakage} \label{sec:syst_I2P}
Mis-estimation of the I2P leakage can produce systematic polarization angle fluctuation.
Since the average polarization fraction of Tau~A at \SI{150}{GHz} is 7\%, the uncertainty of the I2P leakage coefficient of 0.02\% per observation leads to a systematic error of the Tau~A polarization angle of ${\sim}0.1^\circ$ (Eq.~\eqref{eq:I2P_to_syst}).
The systematics due to the uncertainty of I2P leakage was estimated by the difference of Tau~A angle when different models of I2P (Sec.~\ref{sec:cal_I2P}) are applied.
The possible polarization angle fluctuation due to I2P mis-estimation is 0.06$^\circ$ RMS.

\subsection{Ground contamination} \label{sec:syst_ground}
The systematic contamination of data synchronous with the telescope's pointing in ground coordinates is called ground contamination.
During Tau~A observations, the telescope tracks Tau~A by changing azimuth and elevation simultaneously (Sec.~\ref{sec:observations}). 
Due to the complexity of the scan strategy, the subtraction of the ground contamination is not performed, in contrast to the CMB analysis.
Here we evaluate the amount of the ground contamination on the polarization angle of Tau~A.
First, we generate a ground polarization map for each observation by stacking the timestreams in the ground coordinate, masking timestreams around Tau~A within 12$^\prime$ diameter.
A hit map of Tau~A in ground coordinate for each observation is generated simultaneously from the masked timestreams.
The maximum ground polarization signal is found to be about 100~$\mu$K, which is about 200 times smaller than the polarization intensity of Tau~A ($\simeq$20~mK). 
The polarization angle bias due to the ground contamination is evaluated by adding the bias of Stokes parameters $\Delta Q^\mathrm{Ground} = \sum_p^\mathrm{Tau~A~hit}Q_p^\mathrm{Ground}$ and 
$\Delta U^\mathrm{Ground} = \sum_p^\mathrm{Tau~A~hit}U_p^\mathrm{Ground}$ to the Tau~A signal as
\begin{align}
    \psi + \Delta \psi = \frac{1}{2}\tan^{-1} \left(
    \frac{\sum_p^{\diameter<12'} U_p^\mathrm{Tau~A}+\Delta Q^\mathrm{Ground}}
    {\sum_p^{\diameter<12'} Q_p^\mathrm{Tau~A}+\Delta U^\mathrm{Ground}}
    \right).
\end{align}
The possible systematic polarization angle fluctuation due to the ground contamination is 0.03$^\circ$ RMS.

\subsection{Residual 1/f} \label{sec:syst_1f}
The $1/f$ noise also produces systematic polarization angle fluctuation, because the statistical uncertainty for each observation is estimated by the detector-by-detector signflip noise estimation, which cancels the $1/f$ noise, a fraction of which can be common between detectors (Sec.~\ref{sec:map2angle}).
This $1/f$ noise is modeled by stacking all detectors' timestreams in time domain, masking around Tau~A within 12$^\prime$ diameter for each observation.
The stacked timestream is projected to the sky using the pointing of each detector to make a residual $1/f$ map.
The polarization angle bias due to the $1/f$ noise is evaluated by adding the residual $1/f$ noise map to the Tau~A signal as
\begin{align}
    \psi + \Delta \psi = \frac{1}{2}\tan^{-1} \left(
    \frac{\sum_p^{\diameter<12'} U_p^\mathrm{Tau~A}+\sum_p^\mathrm{\diameter<12'}U_p^{1/f}}
    {\sum_p^{\diameter<12'} Q_p^\mathrm{Tau~A}+\sum_p^\mathrm{\diameter<12'}Q_p^{1/f}}
    \right).
    \label{eq:syst_1f}
\end{align}
The systematic polarization angle fluctuation due to the residual $1/f$ is 0.01$^\circ$ RMS.

\subsection{Mizuguchi-Dragone breaking} \label{sec:syst_MD}
The Huan Tran telescope is an off-center Gregorian design which satisfies Mizuguchi-Dragone (MD) condition to cancel the cross-polarization at the primary and secondary mirrors \citep{Mizuguchi1976, Dragone1978}. 
The systematic contamination due to the cross-polarization from breaking the MD condition by the HWP being located at the prime focus of the telescope is expected to be negligibly small.
\textcite{PB_MD_breaking} simulated this effects for the exact same telescope with the \textsc{Polarbear}-2 receiver.
The leading effect of the cross-polarization, the dipole component of $Q$ and $U$ mixing, is considered to put the upper bound on systematics.
The dipole component of the detectors cancel one another due to their focal plane distribution, but the pattern of cancellation differs from day to day, resulting in time-dependent systematic polarization angle fluctuation. 
We simulated the systematic polarization angle fluctuation due to the residual dipole component assuming the worst orientation of the dipole component and 2\% of $Q$ and $U$ mixing. 
The upper bound of the systematic polarization angle fluctuation is 0.01$^\circ$.

\subsection{Pointing} \label{sec:syst_pointing}
The boresight pointing of the telescope propagates to the polarization angle of Tau~A through the transformation of polarization angle from telescope coordinate to sky coordinate.
The observation-by-observation fluctuation of the boresight pointing is evaluated by the variation of the celestial position of the Tau~A intensity signal, which is 0.1$^\prime$ STD and 0.3$^\prime$ STD in right ascension and declination, respectively.
The possible systematic polarization angle fluctuation due to the boresight pointing error is 0.006$^\circ$ RMS.

\subsection{Filtering \& map-making bias} \label{sec:syst_filtering}
Since we adopt the filtered-binned map-making method, the Tau~A angle estimated in map domain is biased (Sec.~\ref{sec:map2angle}).
We evaluated this bias by the full simulation of Tau~A signal timestreams.
The input Tau~A model was constructed by convolving the Tau~A map observed by IRAM \citep{Aumont:2009dx} with the beam of \textsc{Polarbear}.
We simulated timestreams of Tau~A signal scanning the input map, modulating the signal by rotating HWP, and applied the same analysis as the real data to estimate biases.
The systematic polarization angle fluctuation due to the filtering and map-making is 0.005$^\circ$ RMS.

\subsection{Polarization angle of detector} \label{sec:syst_detangle}
The mis-calibration of the polarization angle of detector averaged over the entire observation period can produce systematic polarization angle fluctuation because the data selection is different for each observation. 
The amount of mis-calibration of the polarization angle of each detector, calibrated over the entire observation period, is estimated to be 0.1$^\circ$ RMS, by the variation of the Tau~A polarization angle measured by each detector. 
The possible systematic polarization angle fluctuation is estimated by the average mis-calibration of polarization angle of detectors. 
The estimated systematic polarization angle fluctuation due to the polarization angle of detector is 0.002$^\circ$ RMS.

\section{Discussion} \label{sec:discussion}

Although no point exceeds the significance level of 3$\sigma$, the hint of 2.5$\sigma$ oscillation with an amplitude of 0.11$^\circ$ and 1/61~day$^{-1}$ and 1/52~day$^{-1}$ period deserves further discussion. 
In this section, we enumerate possible causes and describe our investigations.

\subsection{Statistical Uncertainties} \label{sec:stat_error_discussion}
The estimation of statistical uncertainty is verified within the fractional 1$\sigma$ uncertainty of 6.5\% by one of the representative test statistic of the null test, the total $\chi^2_\mathrm{null}$ summed over all frequencies and all data-splits.
The total $\chi^2_\mathrm{null}$ scales inversely proportional to the square of the assumed statistical uncertainty.
We estimated the accuracy of the statistical uncertainty by comparing the measured total $\chi^2_\mathrm{null}$ and the simulated total $\chi^2_\mathrm{null}$.
If we assume uniformly larger statistical errors of 6.5\% for all 95 observations, the significance level of the hint of signal decreases from 2.5$\sigma$ to 2.1$\sigma$. 
Therefore it is possible that the statistical fluctuation of the estimated statistical uncertainty is the cause of the hint of a signal. 

\subsection{Residual systematic errors} \label{sec:residual_systematics}
\begin{figure}
    \centering
    \includegraphics[width = .48\textwidth]{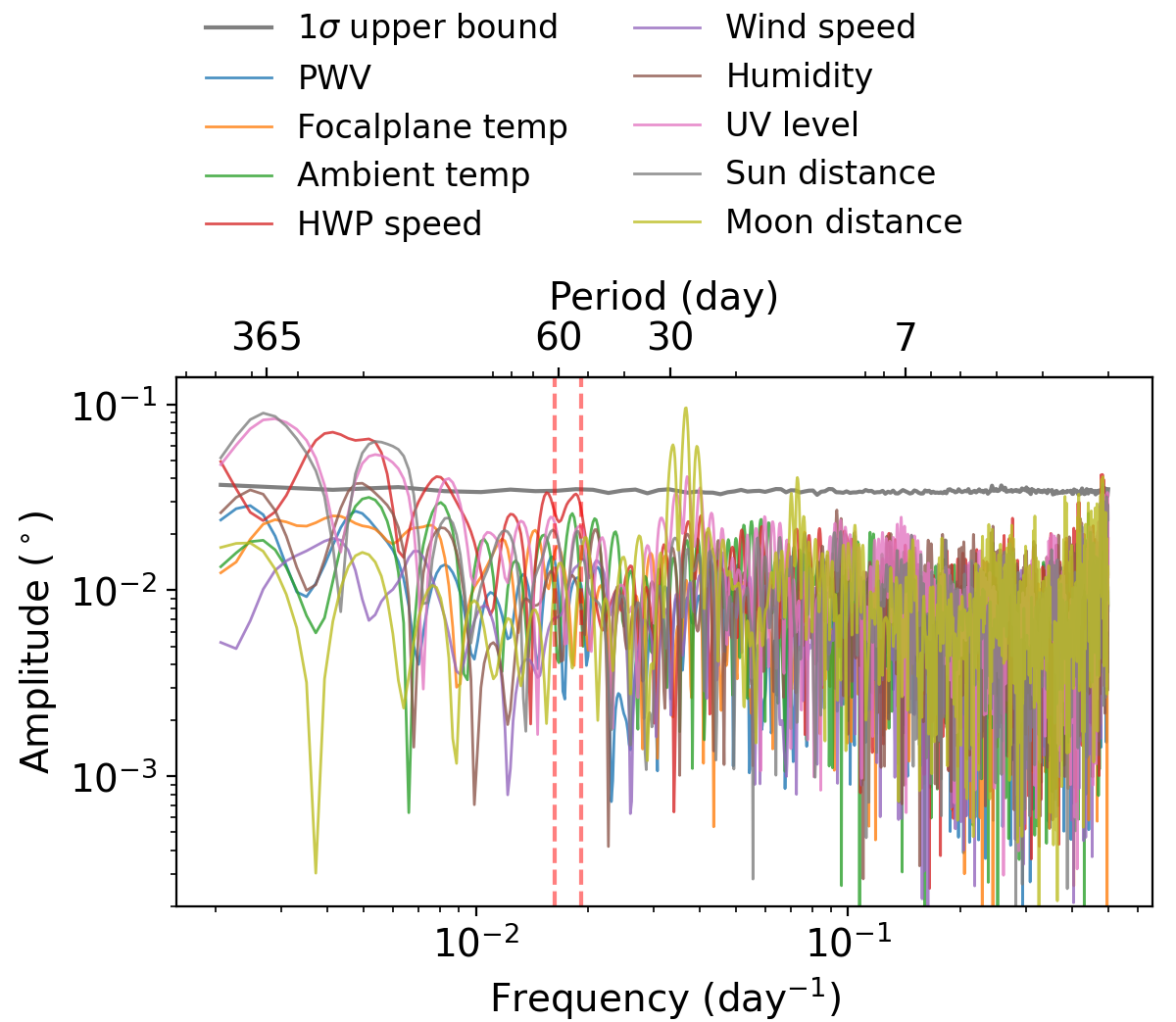}
    \caption{
    2$\sigma$ upperlimit of the LSSA of the systematic templates. 
    The gray solid line shows the 1$\sigma$ upper bound of the statistical uncertainty. The vertical red dashed lines show the frequencies where we found a hint of a signal.}
    \label{fig:corrcoef}
\end{figure}

This section enumerates the investigations of the residual systematic errors. We additionally explore the possible systematic polarization fluctuation using a similar method as Sec~\ref{sec:syst_correlation}. 
We consider additional systematic templates: HWP speed, wind speed, humidity, UV level, Sun distance, and Moon distance, and model the amount of systematic polarization fluctuation in polarization angle units conservatively. 
We model the amount of systematics by assuming that the systematic polarization fluctuation is proportional to the systematic templates. 
The proportionality coefficient ($\hat{c}$) for the conversion of the systematic template into polarization angle units is determined by minimizing the absolute value of the correlation coefficient (Eq.~\eqref{eq:corrcoeff})
\begin{align}
    \hat{c}^\mathrm{best} = \underset{c}{\operatorname{argmin}}\left(|\mathrm{Corrcoef}(T, d-cT)|\right).
    \label{eq:syst_model}
\end{align}
As a conservative estimate, we construct the 95\% upper bound of the proportionality coefficient as
\begin{align}
    &\hat{c}^{95\%} = |\hat{c}^\mathrm{best}| + \mathrm{percentile}\left(\left\{|c|~\middle\vert~c \in Cs\right\}, 95\right)  \\
    &Cs = \left\{\underset{c}{\operatorname{argmin}}\left(|\mathrm{Corrcoef}(T, n_l-cT)|\right)\middle\vert~l=0,1,...,1440\right\},
\end{align}
where $n_l$ is the $l$-th noise realization based on the bootstrap method (Sec.~\ref{sec:map2angle}).
Figure~\ref{fig:corrcoef} shows the spectrum of the possible systematics modeled as $\hat{c}^{95\%}T$ compared with the 1$\sigma$ upper bound of statistical uncertainty. No significant possible systematics are found around the frequency bins where we find the hint of a signal. 
Possible systematics are larger than the 1$\sigma$ upper bound of statistical uncertainty in some frequency bins, however this does not necessarily mean the actual systematics.
This is because these are conservative estimates, and their accuracies are limited by the statistical error of the polarization angle measurements of Tau~A.

\begin{figure}
    \centering
    \includegraphics[width = .49\textwidth]{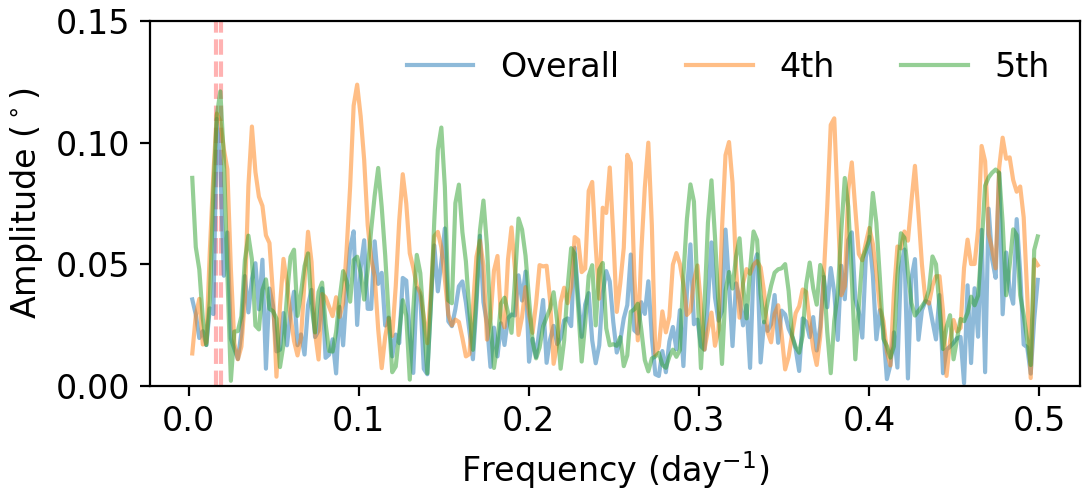}
    \caption{Season by season spectrum of the Tau~A polarization angle. The vertical red dashed lines show the frequencies where we found a hint of a signal. The hint of a signal is consistently seen in two seasons.}
    \label{fig:season_spectrum}
\end{figure}

Another way to probe the systematics is to check the season-by-season consistency of the hint of a signal as shown in Fig.~\ref{fig:season_spectrum}.
The hint of a signal at 1/61\,day$^{-1}$ or 1/52\,day$^{-1}$ is consistently seen in both the 4th and 5th season.
Additionally, we perform the null tests with 10 times higher resolution frequency bins and also by limiting the frequency bins to the two most significant bins. 
No residual systematics with a significance level larger than 2$\sigma$ are found.
Finally, we note that we did not do any on-site maintenance activities on 52 to 61 day cycles.

\begin{figure}
    \centering
    \includegraphics[width = .49\textwidth]{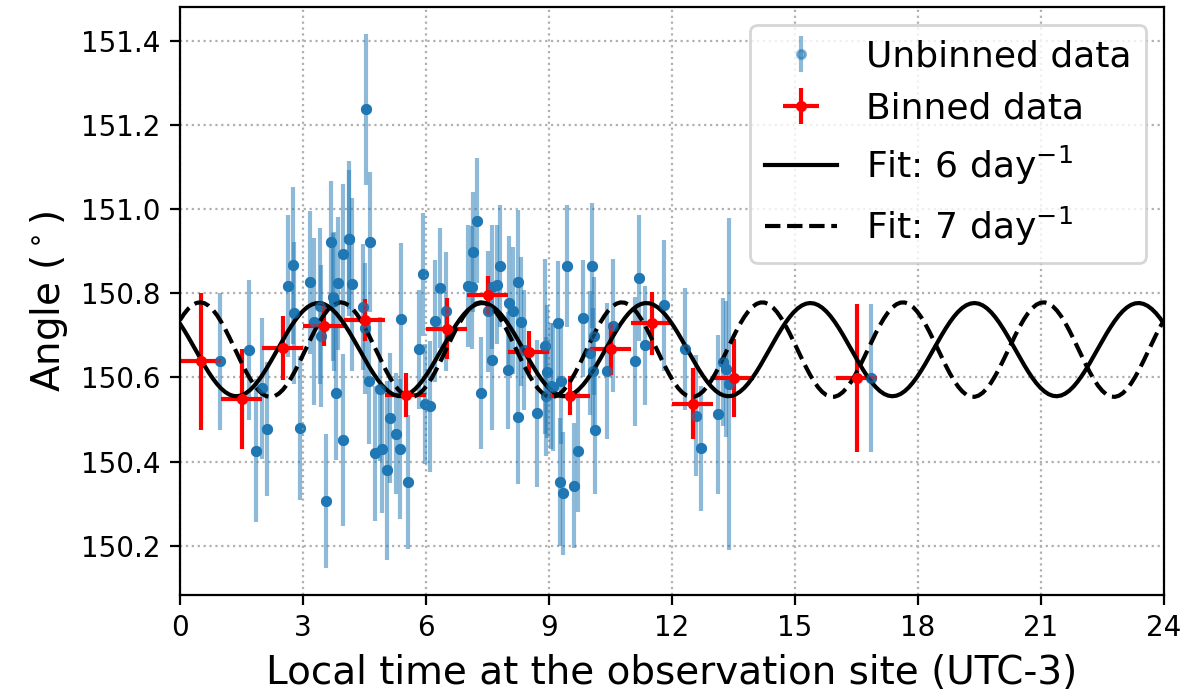}
    \caption{The measured polarization angle of Tau~A as a function of local time at the observation site.}
    \label{fig:localtime_vs_TauA}
\end{figure}

Another possibility of residual systematic error is the aliasing of the diurnal variation of the instrumental systematic error, such as the instrument's rotation around the boresight. 
Figure~\ref{fig:localtime_vs_TauA} shows the measured polarization angle of Tau~A as a function of local time at the observation site. The blue points are the unbinned data, the red points are binned by 1~hour grid and the black solid line and dashed line are the fitted 6\,day$^{-1}$ and 7\,day$^{-1}$ diurnal variation, respectively. 
A possible explanation is that the receiver could be effectively rotated around the boresight by a small amount due to the deformation of the receiver mounting structure caused by the thermal expansion depending on the sunlight exposure, and the measured polarization angle of Tau~A varies diurnally. 
The systematic errors of the boresight pointing due to various environmental conditions including sunlight exposure are evaluated to be smaller than 1$^\prime$ RMS \cite{Matsuda_phd}, which is negligible for this study. Also, only about half of the observed time is directly influenced by the sunlight because the typical sunrise time at the observation site is 7-8~am. 
Nonetheless, we do not have a monitor of the instrument's rotation around the boresight to exclude this possible systematic error. 

\begin{figure}
    \centering
    \includegraphics[width = .49\textwidth]{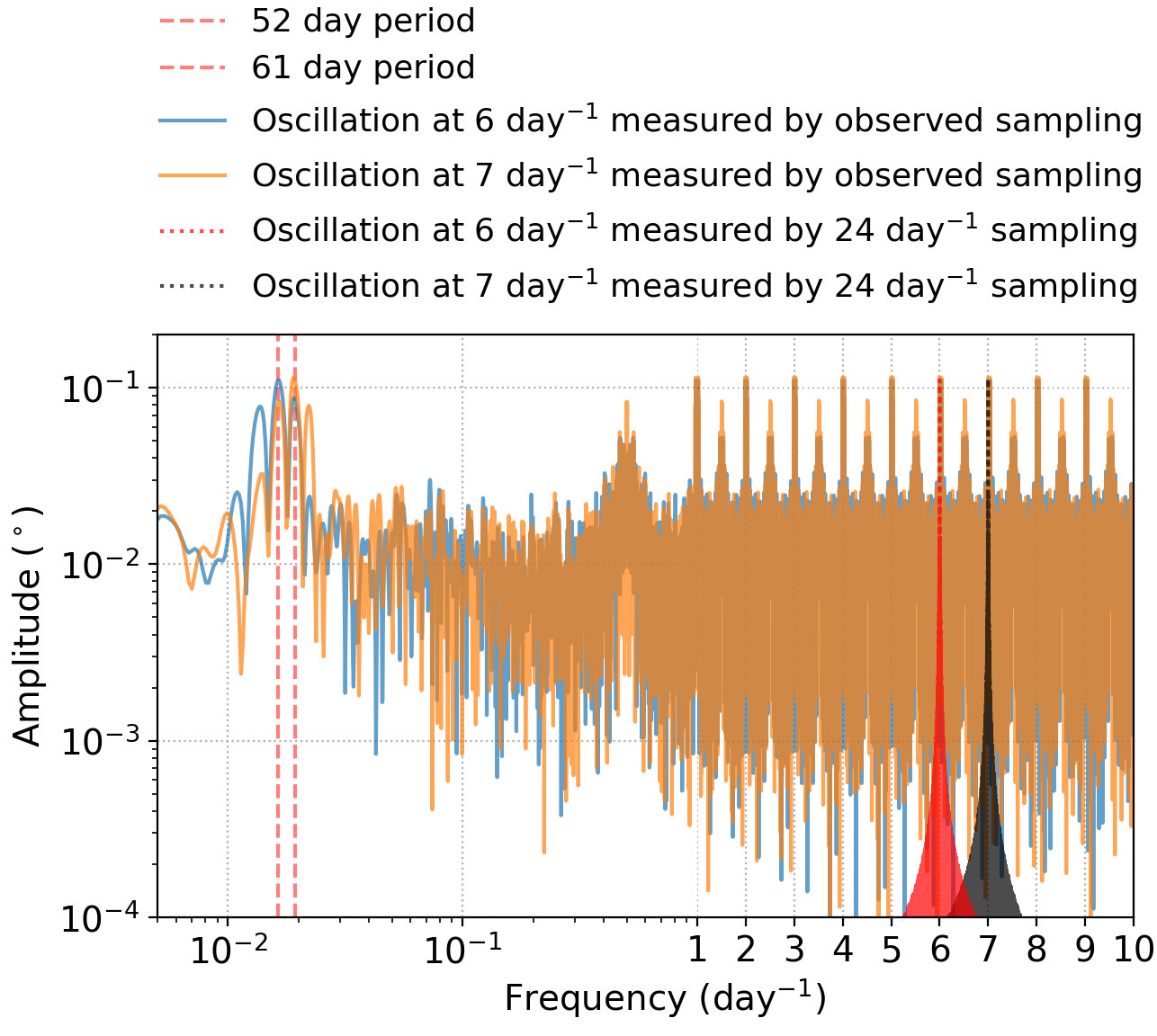}
    \caption{The LSSA of the diurnal variations at 6\,day$^{-1}$ and 7\,day$^{-1}$ measured by the observed sampling schedule. These diurnal variations are aliased to 1/61\,day$^{-1}$ and 1/52\,day$^{-1}$, respectively.}
    \label{fig:aliaising}
\end{figure}

While this may indicate a hint of possible systematic error, it is not statistically significant that the peaks at 1/61\,day$^{-1}$ and 1/52\,day$^{-1}$ can be mapped to 24-hour cycle diurnal signals (Fig.~\ref{fig:aliaising}).
The frequency resolution of our data is set to ~1/486\,day$^{-1}$ by the entire observation period, 486 days. On the other hand, the $n$-th harmonics of 24-hour cycle diurnal variation are aliased to the frequencies of $n/366.24$\,day$^{-1}$ because our observation interval is $1-1/366.24$ day, one local sidereal day.
Thus, there is a high chance of such coincidence for any peak found below $f_\mathrm{NYQ}\simeq0.5$\,day$^{-1}$ with the frequency resolution of 1/486\,day$^{-1}$. 
In our case, $n=6$ and 7 coincides with the 1/61\,day$^{-1}$ and 1/52\,day$^{-1}$, respectively.

We also point out the possibility of the oscillation of the I2P leakage, which can mimic the signal. There are two known mechanisms of creating the I2P leakage. One is due to the optical elements, in which case the I2P leakage tends to be either constant or to gradually increase with time and not to oscillate.  The other is detectors' non-linearity (\citetalias{Takakura:2017ddx}), whose effect would correlate with PWV in our case and thus is investigated in Sec~\ref{sec:syst_correlation}. However, it is worth assessing model-independent constraints we can place on the I2P leakage variability. All the I2P leakage estimation methods discussed in Sec.~\ref{sec:cal_I2P} have different drawbacks to evaluate the time variation of I2P leakage. Method~1 may suffer from time-dependent systematic errors, method~1$^\prime$ assumes no time variation in I2P leakage, and method~2 is under different observation schedule and conditions than Tau A observations. We evaluate the possible systematic polarization fluctuation of Tau~A by adding maximally allowed fluctuations of I2P leakage to the Tau~A signal. According to the calibration uncertainty of method 2, the RMS of possible systematic polarization fluctuation is 0.16$^\circ$ and is not negligible compared to the statistical uncertainty (Table~\ref{table:syst}).
Therefore, we cannot fully exclude this possibility. 


\subsection{Intrinsic variability of Tau~A} \label{sec:variability_taua}
This section explores the possibilities of intrinsic polarization angle variability of Tau~A. Since we measure the average polarization angle of the entire Tau~A, we cannot distinguish the intrinsic variability of Tau~A from the ALP-induced polarization angle oscillation. Tau~A primarily consists of two components, the Crab Nebula and the Crab Pulsar \citep{Hester2008}. The Crab Nebula is an expanding synchrotron nebula. The Crab Pulsar is a point source at the center of the Crab Nebula, the energy source of the Tau~A system. 
Explaining the possible single-mode oscillation seen in our data with the intrinsic variability of these components is not very natural, because of the following two reasons. 
First, the time scale of the single-mode oscillation seen in our data, a period of 52 or 61~days, is not very natural for the Crab Nebula, the expanding synchrotron nebula with the size of several light years. 
Second, the amplitude of the polarization angle oscillation, 0.11 degrees, is too large to be explained by the Crab Pulsar. If the Crab Pulsar had a polarization variability that is \SI{4e-3}{} $= \arctan(2\times0.11\times\pi/180)$ of the Crab Nebula, that could explain the 0.11 degree polarization angle variability of Tau~A. However, at millimeter wavelengths, it is expected to be four orders of magnitude or more darker than the Crab Nebula \cite{Buhler_2014}. Also, a single mode oscillation of the Crab Pulsar with a period around 52 or 61~days is not reported at any wavelength. 
Nevertheless, we explore the possibilities of intrinsic variability of Tau~A using Tau~A measurements in other experiments, because we cannot completely exclude them. 

The time variability of the Tau~A has been studied in a wide variety of timescales and wavelengths. 
To the best of our knowledge, the time variability around 52 or 61~day period has not been observed. 
Here we summarize the known time variation of Tau~A.
\begin{enumerate}
    \setlength{\itemsep}{0pt}
    \setlength{\parskip}{0pt}
    \setlength{\parsep}{0pt}
    \item The Crab Pulsar has a rotational period of about \SI{33}{ms} \citep{Becker:1995bm}.
    \item Giant radio pulses, with durations of $<$ 1\,minute, are observed from Tau~A \citep{Ellingson:2013mta, Eftekhari:2016fyo}.
    \item Variability of the flux at X-ray or Gamma-ray wavelengths from Crab Pulsar exist from years to decades \citep{Kouzu:2013mga, MAGIC:2014izm}.
    \item Large variation in the polarization of Crab Pulsar and Crab Nebula is reported at X-ray wavelengths \citep{Feng:2020miz, Bucciantini2023}.
\end{enumerate}

\begin{figure}
    \centering
    \includegraphics[width = .45\textwidth]{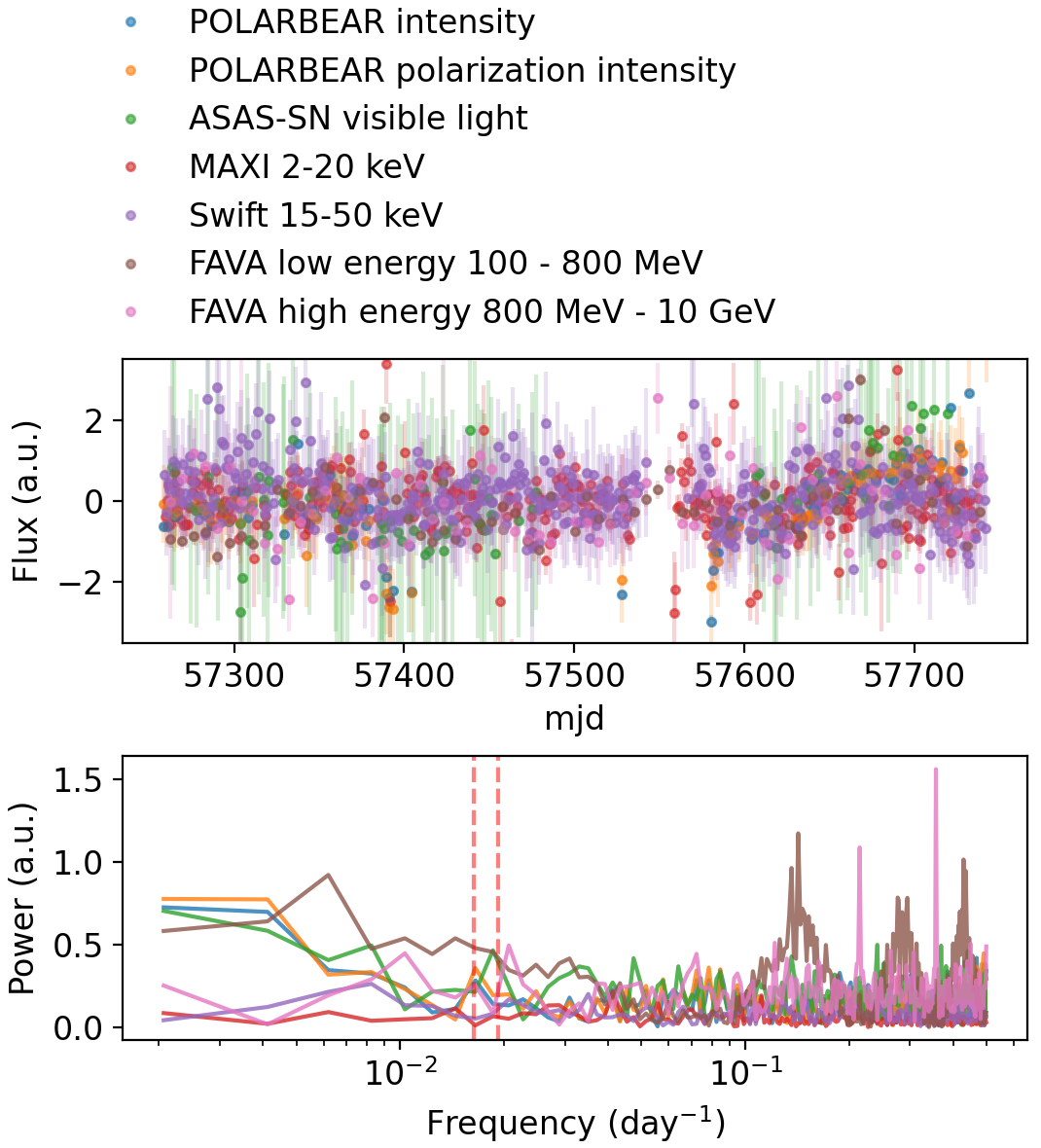}
    \caption{Top: Light-curves of Tau~A measured in this study and various all-sky survey satellites in arbitrary units \cite{Matsuoka:2009nj, Swift2013, Shappee:2013mna, Kochanek:2017wud, Abdollahi:2016rso}. Bottom: LSSA of the light-curves. The vertical red dashed lines show the frequencies where we found a hint of a signal. No clear peaks are found around these frequency bins.}
    \label{fig:satellite_lightcurve}
\end{figure}

The polarization angle of Tau~A has been measured by a number of experiments. However its time dependence has been explored only recently. 
The 95\% upper bound of oscillation amplitude of ${\sim}1^\circ$ was placed by \textcite{Castillo_2022} on periods ranging from \SI{6e-4}{day^{-1}} to \SI{10}{day^{-1}} using data from the QUIJOTE experiment between 10 and \SI{20}{GHz} \citep{Genova-Santos:2015uia}. 
Tau~A has been measured in more experiments for intensity than for polarization.
Therefore, we qualitatively investigate whether the single-mode oscillation seen in our data exists in Tau~A intensity using the public daily light-curve from the all-sky survey satellites over the observation period of this study. 
Figure~\ref{fig:satellite_lightcurve} shows the LSSA of the light-curves measured in this study and various all-sky survey satellites. 
Since these measurements have different cadences than this study, a direct comparison is not possible, although we do not find clear single-mode peaks as seen in our data around the frequency bins where we find the hint of a signal. 


\subsection{Consistency with other ALP bounds} \label{sec:consistency}
\begin{figure}
    \centering
    \includegraphics[width = .45\textwidth]{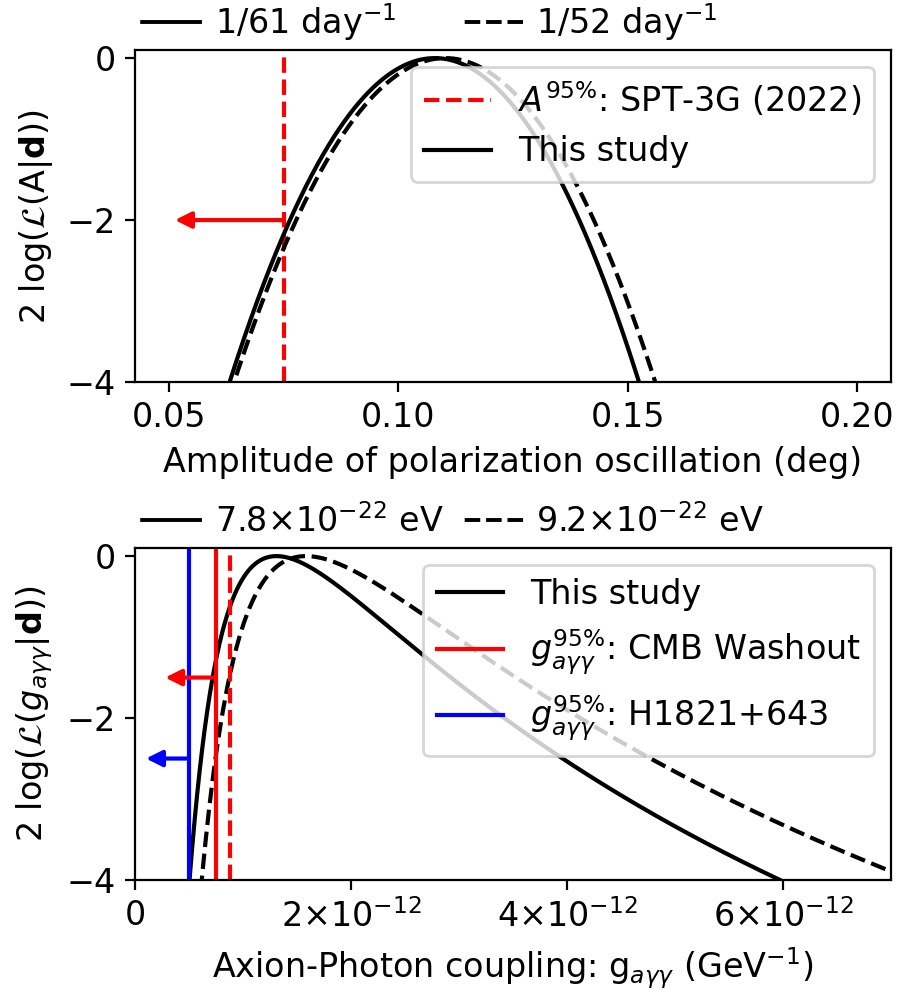}
    \caption{Comparison of the two most significant signals found in this study with other bounds. The red and blue line shows the 95\% upper bound by other measurements and the black line shows the likelihood of this study. Top panel compares the amplitude of an axion-like polarization oscillation. Bottom panel compares the ALP-photon coupling.}
    \label{fig:consistency}
\end{figure}

We discuss the consistency of our results with other ALP bounds. 
The consistency analysis discussed in this section is only valid if the hint of a signal we found is purely from ALP.
To the best of our knowledge, the second tightest upper bound of an axion-like polarization oscillation of a local ALP field is placed by \textcite{ALP_SPT3g}, with the median upper bound of 0.071$^\circ$ from 1/100\,day$^{-1}$ to 1\,day$^{-1}$, while our median upper bound is 0.065$^\circ$. 
The largest discrepancy is found at the frequencies where we found the hint of oscillation. 
Fig~\ref{fig:consistency} indicates the significance of the discrepancy.

A more significant discrepancy is found when we translate the polarization oscillation to a constraint on ALP-photon coupling.
One of the tightest upper bounds of the axion-photon coupling is placed by the absence of the suppression of CMB $E$-mode power spectrum due to an axion-like oscillation in the recombination era \citep{Fedderke2019}. 
\begin{align}
    g_{a\gamma\gamma} \leq (\SI{9.6e-13}{GeV^{-1}}) \times\left(\frac{m_a}{10^{-21}\,\mathrm{eV}}\right).
\end{align}
The 95.5\% upper bound of \SI{5.0e-13}{GeV^{-1}} at $m_a < $\SI{10e-12}{eV} is also placed from the absence of the distortions of the X-ray spectrum of the quasar H1821+643 due to the ALP-photon coupling \citep{Chandra2022}. 
The bottom panel of Fig.~\ref{fig:consistency} indicates the significance of the discrepancy. 
Our bounds are complementary to these bounds because the ALP search by an axion-like oscillation is less dependent on the cosmological model or the model for the galaxy cluster magnetic fields.
 
Several analyses have claimed to rule out the ultralight dark matter in mass region where we found a hint of a signal, $m_a < \SI{2.0e-21}{eV}$ by \cite{lyman_alpha_fuzzy} and $m_a < \SI{2.9e-21}{eV}$ by \cite{Nadler2021}.
Since there are model dependencies in these analyses, our analysis is complementary in searching for an entirely different signal. Our analysis applies even if the ALP constitutes a fraction of the dark matter, the bounds are inversely proportional to the square root of the dark matter fraction, $\kappa$. 

We also check the consistency with our previous result, the search for the polarization oscillation in the CMB \citetalias{ALP_PB}. 
The hint of a signal we find is below the noise level of \citetalias{ALP_PB}, and \citetalias{ALP_PB} is consistent with this study.

\section{Conclusion} \label{sec:conclusion}
We analyze the Tau~A observations in the 4th and 5th seasons of the \textsc{Polarbear} experiment, and present a median 95\% upper bound of $A < 0.065^\circ$ on the polarization oscillation amplitude of Tau~A over oscillation frequencies from \SI{0.75}{year^{-1}} to \SI{0.66}{hour^{-1}}.
To the best of our knowledge, this work provides the most accurate measurements of the amplitude of the polarization oscillation of Tau~A. 
Our improved analysis of Tau~A data introduces new methodologies for investigating possible systematics, which allowed us to improve the polarization angle calibration method, and demonstrate a sensitive search for an axion-like polarization oscillation using the calibration data from the \textsc{Polarbear} ground-based CMB observatory located in the Atacama Desert of Chile. 
Our measurements constrain potential intrinsic time variability in Tau~A, which is an important understanding of Tau~A as a polarization calibrator, and may be used to better understand the dynamics of Tau~A. 
Under the assumption that no sources other than an ALP are causing Tau A’s polarization angle variation, that the ALP constitutes all the dark matter, and that the ALP field is a stochastic Gaussian field, this bound translates into a median 95\% upper bound of an ALP-photon coupling $g_{a\gamma\gamma} < 2.16\times10^{-12}\,\mathrm{GeV}^{-1}\times(m_a/10^{-21} \mathrm{eV})$ in the mass range from \SI{9.9e-23}{eV} to \SI{7.7e-19}{eV}. 
To the best of our knowledge this is the tightest bound from the measurement of the ALP-induced polarization angle oscillation. This study demonstrates this type of analysis using bright polarized sources are competitive as those using the polarization of CMB in constraining ALPs. 
Analysis using the polarization of CMB has the advantage of being immune to the complexity of the dynamics of the sources. Using multiple polarized sources and cross-correlating their time variations will lead to a more robust analysis. 


We find a hint of a polarization oscillation signal with amplitude $0.11^\circ$ at 1/61\,day$^{-1}$ with a global significance level of 2.5$\sigma$, and an equally significant correlated polarization oscillation at 1/52\,day$^{-1}$. 
We consider whether this hint of a signal is due to residual systematic errors or intrinsic variability of Tau~A due to yet unknown dynamics, but so far we find no clear evidence for either of them.  
There are possible residual systematic errors we do not exclude.
Interpreting this as an ALP signal leads to significant tension with other constraints.
We expect that future measurements with more data will not only reduce statistical errors but also allow us to better understand possible systematic errors, leading us to a more conclusive interpretation.
A successor experiment, the Simons Array, employs dichroic detectors including \SI{90}{GHz} and \SI{150}{GHz} center band \citep{POLARBEAR:2015ixw}.
Since the polarized intensity of the synchrotron radiation is proportional to $\nu^{-0.3}$ \citep{Weiland:2010ij}, we expect a more sensitive observation from the lower frequency band.  
Also, since an axion-like polarization oscillation is independent of the optical frequency of light, the control of frequency-dependent systematic errors is possible.



\begin{acknowledgments}
The \textsc{Polarbear} project is funded by the National Science Foundation under grants AST-0618398 and AST-1212230. 
This work is supported in part by the Gordon and Betty Moore Foundation grant number GBMF7939, and by the Simons Foundation Grant Number 034079. 
In Japan, this work is supported by JSPS KAKENHI Grant Number 18H05539, 19H00674, 22H04945, 23H00105 and JSPS Core-to-Core program Grant Number JPJSCCA20200003, and World Premier International Research Center Initiative (WPI), MEXT, Japan. 
Work at LBNL is supported in part by the U.S. Department of Energy, Office of Science, Office of High Energy Physics under contract No. DE-AC0205CH11231.  
Calculations were performed on the Central Computing System, owned and operated by the Computing Research Center at KEK, and the National Energy Research Scientiﬁc Computing Center, which is supported by the Department of Energy under Contract No.DE-AC02-05CH11231. 
The UC San Diego group acknowledges support from the James B. Ax Family Foundation. 
The Melbourne group acknowledges support from the Australian Research Council's Discovery Projects scheme (DP210102386). 
C.B. acknowledges support by the HORIZON-CL4-2023-SPACE-01, GA 101135036 RadioForegroundsPlus Project, by the COSMOS and LiteBIRD networks of the Italian Space Agency, and by the InDark and LiteBIRD Initiative of the National Institute for Nuclear Physics (INFN).  
Y.C. acknowledges the support from JSPS KAKENHI Grant Number 21K03585. 
J.E. acknowledges the SCIPOL project funded by the European Research Council (ERC) under the European Union’s Horizon 2020 research and innovation program (Grant agreement No. 101044073)
S.T. acknowledges JSPS Overseas Research Fellowships and KAKENHI Grant Number JP14J01662, JP18J02133, JP18H04362, and JP20K14481.
K.Y. acknowledges the support from JSPS KAKENHI Grant Number 21J11179, and the support from XPS, WINGS Programs, The University of Tokyo. 
This research has made use of the data provided by the IRAM 30 meter telescope, 
the MAXI data provided by RIKEN, JAXA and the MAXI team, 
the Swift/BAT transient monitor results provided by the Swift/BAT team,
the ASAS-SN light-curves provided by the ASAS-SN Team,
the Fermi All-sky Variability Analysis provided by the Fermi-LAT Collaboration.
We thank Aya Bamba, Adriaan Duivenvoorden, Gregory Gabadadze, Takeo Moroi, Lyman Page, and David Spergel for useful discussions.

\end{acknowledgments}

\appendix
\section{Transfer function} \label{apx:transfer_function}
This section describes the derivation of the transfer functions shown in Fig.~\ref{fig:transferfunction}.
First, the estimation of Tau~A angle by averaging over the single observation period of one hour biases the oscillation signal at higher frequency. We consider observing an axion-like polarization oscillation of Tau~A at $f$ with unit amplitude and phase $\theta$ as
\begin{align}
    \psi(t) = \psi_0 + \sin(2\pi f t +\theta).
    \label{eq:input}
\end{align}
The polarization oscillation averaged from $t_s$ to $t_e$ is
\begin{align}
    \psi_t 
    &= \frac{1}{t_e-t_s}\int_{t_s}^{t_e} dt'~\psi(t') \\
    &=  \psi_0 + \mathrm{sinc}(\pi f T_\mathrm{obs})\sin\left(2\pi f t+\theta\right),
\end{align}
where $T_\mathrm{obs} = t_e-t_s (\simeq \mathrm{one~hour})$ is the observation period, $t=(t_s+t_e)/2$ is the average observation time, and we used a unnormalized sinc function. 
The corresponding transfer function is 
\begin{align}
    F_f^\mathrm{average} = \mathrm{sinc}(\pi f T_\mathrm{obs}).
    \label{eq:transfer_ave}
\end{align}
Second, the observation timing jitter acts like a low-pass filter, but is negligible compared to the time averaging. 

Third, the calibration of the relative polarization angle between detectors (Sec.~\ref{sec:polcaldet}) also biases the signal. With an axion-like polarization oscillation of Eq.~\eqref{eq:input}, the polarization angle of Tau~A measured by the $i$-th detector will be biased by
\begin{align}
    \Delta \psi_i = \sum_{t\in t(i)} (\psi_t \sigma^{-2}_{t,i})/\sum_{t\in t(i)}\sigma^{-2}_{t,i} -\psi_0,
    \label{eq:each_det_bias}
\end{align}
where $\sigma_{t,i}$ is the $i$-th detector's statistical uncertainty of the polarization angle of Tau~A per observation, and $t(i)$ is timing of the set of observations when the $i$-th detector is operating. 
Since we calibrate each detector's polarization angle using the Tau~A polarization angle averaged over the entire observation period, $\Delta \psi_i$ is equivalent to the amount of each detector's mis-calibration.
Then, the measured axion-like polarization oscillation is biased as 
\begin{align}
    \hat{\psi}_t &= \psi_t - \frac{\sum_{i\in i(t)}  \Delta \psi_i \sigma_{t,i}^{-2}}{\sum_{i\in i(t)}\sigma^{-2}_{t,i}},
     \label{eq:transfer_detcal}
\end{align}
where $i(t)$ is a set of detectors working in the observation performed at $t$. 
We calculate the bias from the mis-calibration of the polarization angle of detectors by averaging over the phase of oscillation for each frequency as 
\begin{align}
    F_f^\mathrm{polcal} = \frac{1}{2\pi}\int_0^{2\pi} \left\vert\sum_t F_{ft} \hat{\psi}_t\right\vert d\theta.
\end{align}
Finally, the calibration of the absolute polarization angle ($\psi_0$) also biases the LSSA as
\begin{align}
    F_f^\mathrm{abscal} = \frac{1}{2\pi}\int_0^{2\pi} \left\vert\sum_t F_{ft} \left(\hat{\psi}_t-
    \frac{\sum_t \hat{\psi}_t\sigma_t^{-2}}{\sum_t\sigma_t^{-2}} \right)\right\vert d\theta.
\end{align}
The overall transfer function $F_f$ is evaluated by simulating all the effects at once. The deviation of $F_f$ from unity primary comes from $F^\mathrm{average}_f$, and for the smoothed results (Eqs.~\eqref{eq:median_A}, \eqref{eq:median_stoch} and \eqref{eq:median_det}) we approximate the overall transfer function $F_f$ as
\begin{align}
    F_f^\mathrm{smoothed} \simeq F^\mathrm{average}_f\times\braket{F_f}_{f<f_\mathrm{NYQ}}. 
    \label{eq:smoothed_transfer}
\end{align}

\section{Estimator} \label{apx:estimator}
This section compares the methods for estimating $g_{a\gamma\gamma}$ used in this study and \citetalias{ALP_PB}.
In this study, we estimate $g_{a\gamma\gamma}$ by maximizing the probability density for obtaining the LSSA of the data. We call this estimator the ``frequency estimator."  
This is motivated by considering a simplified case where the noise covariance matrix (Eq.~\eqref{eq:ncov}) is diagonal, i.e., when the timestamp is regularly spaced and the error bars are uniformly distributed. 
In this case, $2|\tilde{d}_f|^2/\braket{|n_f|^2}$ follows chi-square distribution with two degrees of freedom and Eq.~\eqref{eq:P_df_g} is simplified as 
\begin{align}
    P(\tilde{d}_f|g_{a\gamma\gamma}, \phi_\mathrm{DM}) 
    &= \frac{\exp\left(-\frac{|\tilde{d}_f|^2}{\braket{|\tilde{n}_f|^2} + g_{a\gamma\gamma}^2 F_f^2 \phi_\mathrm{DM}^2/4}\right)}{\pi(\braket{|\tilde{n}_f|^2} + g_{a\gamma\gamma}^2 F_f^2 \phi_\mathrm{DM}^2/4)}.
\end{align}
Eq.~\eqref{eq:g2estimator} is a maximum likelihood estimator for this likelihood function. 

On the other hand, \citetalias{ALP_PB} estimates the $g_{a\gamma\gamma}$ by maximizing the probability density for obtaining time-domain data. 
We call this estimator the ``full estimator." The likelihood function is 
\begin{align}
    &P(d_t|g_{a\gamma\gamma}) = 
    \iint d\phi_c d\phi_s~P(d_t|\phi_c, \phi_s, g_{a\gamma\gamma})P(\phi_c)P(\phi_s),
    \label{eq:full_likelihood}
\end{align}
where 
\begin{align}
    P(\phi_c) = \frac{\exp\left(-\frac{\phi_c^2}{\phi_\mathrm{DM}^2}\right)}{\sqrt{\pi\phi_\mathrm{DM}^2}},~
    P(\phi_s) = \frac{\exp\left(-\frac{\phi_s^2}{\phi_\mathrm{DM}^2}\right)}{\sqrt{\pi\phi_\mathrm{DM}^2}}
\end{align}
is the probability density for the stochastic oscillation amplitudes of ALP, and
\begin{align}
    &P(d_t|\phi_c, \phi_s, g_{a\gamma\gamma}) = \prod_t\frac{e^{-\frac{1}{2}\left[(d_t-\psi(t))/\sigma_t\right]^2}}{\sqrt{2\pi\sigma_t^2}}
\end{align}
is the probability density for obtaining data when the stochastic oscillation amplitudes of ALP are given.

\begin{figure}
    \centering
    \includegraphics[width = .49\textwidth]{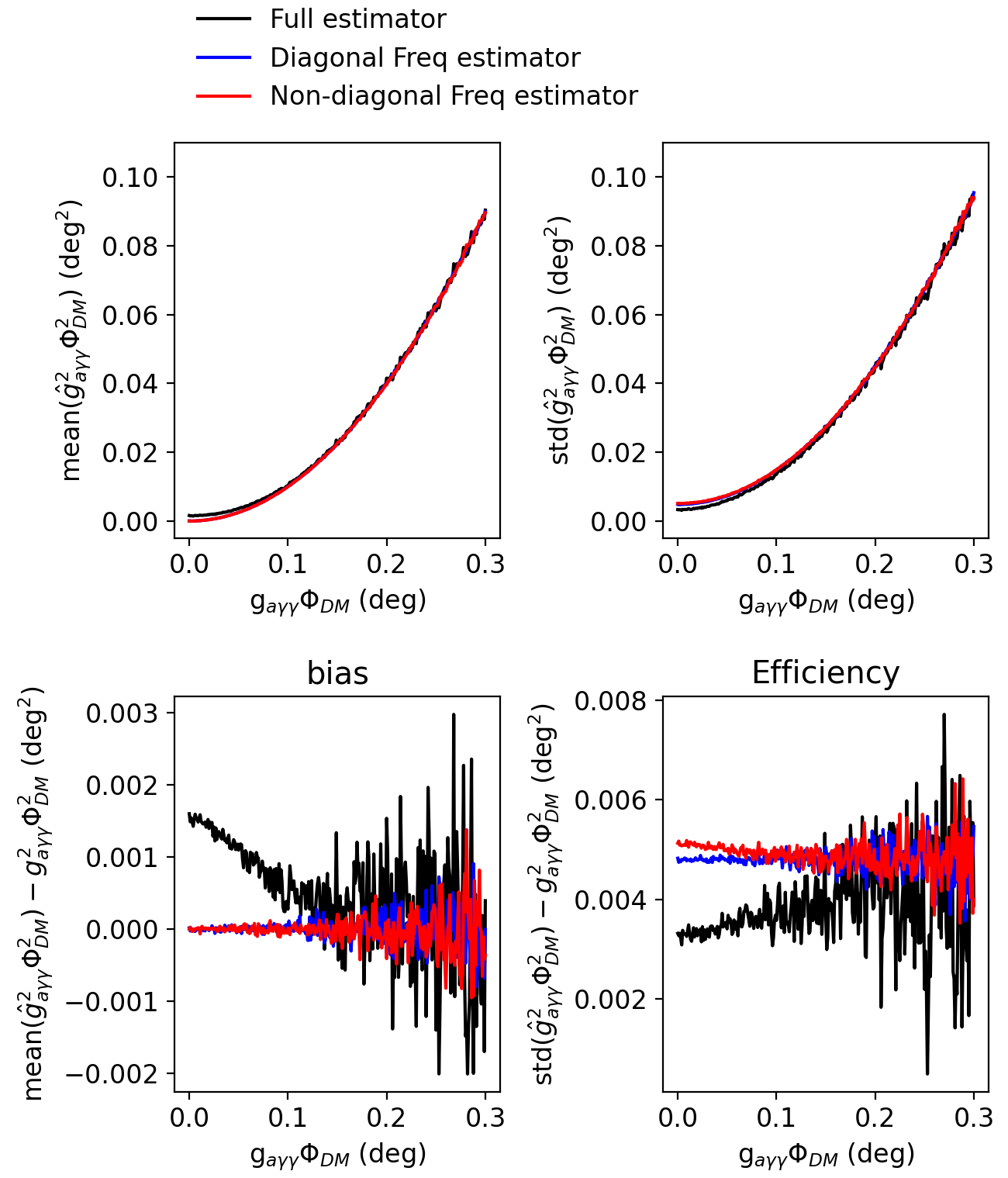}
    \caption{Comparison of the bias and efficiency of the estimators of $g_\mathrm{a\gamma\gamma}^2$. 
    The typical non-diagonality of our noise covariance matrix ($2\tilde{N}_f^{01}/\mathrm{trace}(\tilde{N}_f)$, $
    (\tilde{N}_f^{00}-\tilde{N}_f^{11})/\mathrm{trace}(\tilde{N}_f))$ is 6\%. 
    The estimator for the frequency bin with the largest non-diagonality of 33\% is compared.}
    \label{fig:bias_efficiency}
\end{figure}

Figure~\ref{fig:bias_efficiency} compares the bias and efficiency of the ``frequency estimator" and ``full estimator." 
The ``full estimator" is biased when the signal is small. 
The ``frequency estimator" is a good unbiased estimator regardless of whether the noise covariance is diagonal or not, and the efficiency is slightly larger than ``full estimator."  
Another advantage of the ``frequency estimator" is its small computational cost.
This is because ``full estimator" requires convolution for constructing likelihood (Eq.~\eqref{eq:full_likelihood}), while the corresponding convolution is analytically conducted for the ``frequency estimator" (Sec.~\ref{sec:estimate_g}).

\section{Upper bound}  \label{apx:bound}
This section describes the method for estimating the upper bound of $g_{a\gamma\gamma}$. 
First, we consider the constraint on $g^2\equiv g_{a\gamma\gamma}^2 F_f^2 \phi^2_\mathrm{DM}/4$, the square of the effective amplitude of polarization oscillation, then we convert it to a constraint on $g_{a\gamma\gamma}$.
The probability density for $\hat{g}^2\equiv \hat{g}_{a\gamma\gamma}^2 F_f^2 \phi^2_\mathrm{DM}/4$ given $g^2_\mathrm{true}$, $P(\hat{g}^2|g^2_\mathrm{true})$, can be generated by Monte Carlo simulations using Eqs.~\eqref{eq:g2estimator} and \eqref{eq:P_df_g}.
To quantify the preference of the null-hypothesis over the presence of signal, 
we consider the ratio between the probability of null-hypothesis $P_\mathrm{null}$ and the alternative hypothesis $P_\mathrm{alt}$. 
The probability $P_\mathrm{null}$ is obtained as
\begin{align}
    &P_\mathrm{null}(g^2_\mathrm{true}) 
    = \int_{\Sigma(\hat{g}^2)} P(\hat{g}^2|g^2_\mathrm{true})\,d\hat{g}^2,
\end{align}
where the integration range is determined to satisfy following three conditions:
\begin{align}
    &\Sigma(\hat{g}^2) = \set{\hat{g}^2 | \hat{g}^2 \leq \hat{g}^2_1~\mathrm{or}~\hat{g}^2_2 \leq \hat{g}^2} \\
    & \int_{-\infty}^{\hat{g}_1^2} P(\hat{g}^2 |g^2_\mathrm{true})\,d\hat{g}^2 = \int_{\hat{g}_2^2}^{\infty} P(\hat{g}_2^2 | g^2_\mathrm{true})\,d\hat{g}^2 \\
    & \hat{g}_1^2 = \hat{g}^2_\mathrm{obs}~\mathrm{or}~\hat{g}_2^2 = \hat{g}^2_\mathrm{obs}.
\end{align}
The probability $P_\mathrm{alt}$ is obtained in similar way, $P_\mathrm{alt} = P_\mathrm{null}(g^2_\mathrm{best}(\hat{g}^2_\mathrm{obs}))$. The $g^2_\mathrm{best}(\hat{g}^2_\mathrm{obs})$ is $g^2_\mathrm{true}$ which maximizes $P_\mathrm{null}$ within the physical region.
Finally, we obtain the 90\% confidence interval of $g^2$ as
\begin{align}
    \mathrm{CI} = \left\{g^2_\mathrm{true}~\middle\vert~\frac{P_\mathrm{null}(g^2_\mathrm{true})}{P_\mathrm{alt}} \geq 0.9 \right\}
\end{align}
Since we do not find evidence for a signal with significance level larger than 3$\sigma$, we only report a 95\% upper bound from the upper side of the 90\% confidence interval.
The upper bound of $g_{a\gamma\gamma}F_f\phi_\mathrm{DM}/2$ is square root of that of $g^2$. 
From the relation between the mean amplitude of ALP ($=\phi_\mathrm{DM}$) and the dark matter density (Eq.~\eqref{eq:phiDM}), the 95\% upper bound on $g_{a\gamma\gamma}$ is 
\begin{align}
    g_{a\gamma\gamma} \leq &(\SI{1.6e-11}{GeV^{-1}}) \nonumber \\
    &\times\left(\frac{\sqrt{g^{2,95\%}}/F_f}{1^\circ}\right)\left(\frac{m_a}{10^{-21}\,\mathrm{eV}}\right)
    \left(\frac{\kappa\rho_0}{0.3\,\mathrm{GeV/cm^3}}\right)^{-1/2},
    \label{eq:upperbound}
\end{align}
where $\kappa$ is the fraction of the dark matter that ALP constitutes, and $\rho_0$ is the local dark matter density.


%

\end{document}